\documentclass[aps,preprintnumbers,superscriptaddress,longbibliography]{revtex4-1}
\usepackage{amsmath, mathrsfs, amssymb,amsfonts,amsthm,graphicx, epsf, dcolumn, yfonts}
\usepackage[normalem]{ulem}
\usepackage{xcolor}
\usepackage{amssymb}
\usepackage{slashed}
\usepackage{setspace}
\usepackage{cancel}
\usepackage{wasysym}
\usepackage{tikz-feynman}
\usepackage{float}
\usepackage[utf8]{inputenc}
\usepackage{todonotes}
\usepackage{tikz}
\usepackage{tikz-cd}
\usepackage{units}
\usepackage{subcaption}
\usepackage{array}
\usepackage{tabularray}
\usepackage{tabularx}
\usepackage{physics}
\definecolor{subduedgreen}{RGB}{0,215,0} 
\definecolor{subduedblue}{RGB}{0,0,255} 
\usepackage[hyperfootnotes=true]{hyperref}
\hypersetup{
    breaklinks=true,
    colorlinks=true,
    citecolor=subduedgreen, 
    linkcolor=subduedblue, 
} 
\pdfoutput=1
\newcommand{\lin}{L}

\newcommand{\rind}{+}
\newcommand{\lind}{-}
\newcommand{\q}{q}

\newcommand{\xq}{\phi}

\newcommand{\xql}{\phi^-}
\newcommand{\xqr}{\phi^+}

\def\<{\langle}
\def\>{\rangle}

\newcommand{\cqstate}{\varrho}

\newcommand{\be}{\begin{eqnarray} \begin{aligned}}
\newcommand{\ee}{\end{aligned} \end{eqnarray} }
\newcommand{\benn}{\begin{eqnarray*} \begin{aligned}}
\newcommand{\eenn}{\end{aligned} \end{eqnarray*} }

\newcommand{\ben}{\begin{eqnarray} \begin{aligned}}
\newcommand{\een}{\end{aligned} \end{eqnarray} }

\newcommand{\bc}{\begin{center}}
\newcommand{\ec}{\end{center}}

\newcommand{\id}{\mathbb{1}}

\newcommand{\beq}{\begin{eqnarray} \begin{aligned}}
\newcommand{\eeq}{\end{aligned} \end{eqnarray} }
\newcommand{\bea}{\begin{array}}
\newcommand{\eea}{\end{array}}

\newcommand{\bee}{\begin{enumerate}}
\newcommand{\eee}{\end{enumerate}}
\newcommand{\bei}{\begin{itemize}}
\newcommand{\eei}{\end{itemize}}

\usepackage{amsfonts}

\def\01{\{0,1\}}

\newcommand{\cE}{\mathcal{E}}

\newcommand{\str}{{\vec{x}}}

\def\<{\langle}
\def\>{\rangle}

\def\s{\,\,\,\,}

\newcommand{\sgn}{\operatorname{sgn}}

\def\munu{{\mu\nu}}

\newtheorem*{rep@theorem}{\rep@title}
\newcommand{\newreptheorem}[2]{%
\newenvironment{rep#1}[1]{%
 \def\rep@title{#2 \ref{##1} (restatement)}%
 \begin{rep@theorem}}%
 {\end{rep@theorem}}}
\makeatother

\newreptheorem{thm}{Theorem}
\newreptheorem{lem}{Lemma}

\def\T{\bf T}

\def\z{{q,\dot{q}}}

\def\g{{g}}

\def\T00{{\bf T_{NN}}}

\def\0mom{{\bar{\Gamma}^{\alpha\beta}(\z)}}
\def\1mom{{\Gamma^{\alpha\beta}_1(\z)}}
\def\2mom{{\Gamma^{\alpha\beta}_2(\z)}}

\def\R2Term{\Delta_{\text{\tiny 2}}}
\def\WeylTerm{\Delta_{\text{\tiny w}}}
\begin{document}

\title{Renormalisation of postquantum-classical gravity}

\author{Andrzej Grudka}
\affiliation{Institute of Spintronics and Quantum Information, Faculty of Physics and Astronomy, Adam Mickiewicz University, 61-614 Poznań, Poland}

\author{Tim R. Morris}
\affiliation{STAG Research Centre, Department of Physics and Astronomy,
University of Southampton, Highfield, Southampton, SO17 1BJ, United Kingdom}
\author{Jonathan Oppenheim}

\affiliation{Department of Physics and Astronomy, University College London, Gower Street, London WC1E 6BT, United Kingdom}
\author{Andrea Russo}
\affiliation{Department of Physics and Astronomy, University College London, Gower Street, London WC1E 6BT, United Kingdom}

\author{Muhammad Sajjad}
\affiliation{Department of Physics, Middle East Technical University, 06800, Ankara, Turkiye}

\begin{abstract}
One of the obstacles to reconciling quantum theory with general relativity, is constructing a theory which is both consistent with observation, and and gives finite answers at high energy, so that the theory holds at arbitrarily short distances. Quantum field theory achieves this through the process of renormalisation, but famously, perturbative quantum gravity fails to be renormalisable, even without coupling to matter.  
Recently, an alternative to quantum gravity has been proposed, in which the geometry of spacetime is taken to be classical rather than quantum, while still being coupled to quantum matter fields~\cite{oppenheim2018post,oppenheim2023covariant}. 
This can be done consistently, provided the dynamics is fundamentally stochastic. Here, we find that the pure gravity theory is formally renormalisable. We do so via the path integral formulation by relating the classical-quantum action to that of quadratic gravity. Because the action induces stochastic dynamics of spacetime, rather than deterministic evolution of a quantum field, the classical-quantum theory is free of tachyons and negative norm ghosts. The key remaining question is whether the renormalisation prescription retains completely positive (CP) dynamics. This consideration appears to single out
the scale invariant and asymptotically free theory. We give further evidence that the theory is CP, by showing that the two-point function of the scalar mode is positive semidefinate.  To support the use of precision accelerometers in testing the quantum nature of spacetime, we compute the power spectral density of the acceleration. The variance in acceleration is determined by a dimensionless coupling constant which can be bound via tabletop coherence experiments. We find that relativistic corrections significantly alter the two-point function of the gravitational field, leading to milder stochastic fluctuations. We compute the corrected acceleration spectral density and show that current upper and lower experimental bounds are sufficiently tight to enable tabletop tests of the quantum versus classical nature of spacetime in the near term. With a fuller understanding of the dynamics it may also enable tests on an astrophysical scale. 
\end{abstract}

\maketitle

\section{Introduction}
\label{sec: intro}

In Einstein's theory of general relativity, matter causes spacetime to curve, via 
\begin{equation}
\label{eq:einstein}
G^{\mu\nu}+\Lambda g^{\mu\nu}=\frac{8\pi G_N}{c^4}T^{\mu\nu}
\end{equation}
where the left-hand side encodes the spacetime degrees of freedom via the Einstein tensor $G^{\mu\nu}$, and on the right-hand side sits the energy-momentum tensor $T^{\mu\nu}$ encoding the matter degrees of freedom. Since matter has a quantum nature, the right-hand side of Equation~\eqref{eq:einstein}  must become an operator $\hat{T}^{\mu\nu}$, according to quantum theory (we will henceforth denote quantum operators by placing a $\hat{}$ on them, use units where $\hbar=c=1$, and take the metric signature to be $-,+,+,+$). Consistency would then appear to require that the Einstein tensor become an operator as well, and there has been virtual unanimity in pursuing this approach of quantising gravity. However, the initial difficulty with constructing a quantum theory of gravity, arose because perturbative quantum gravity fails to be renormalisable. Over-coming this issue is one of the central challenges in reconciling quantum theory with gravity. Loop quantum gravity~\cite{sen1982gravity,ashtekar1986new,rovelli1988knot,rovelli2004quantum} achieves UV-finiteness, because the spin-foam network degrees of freedom essentially live on a lattice, which introduces a cut-off related to the spacing of the lattice. String theory~\cite{veneziano1968construction,polchinski2005string,Green:1987sp,green2012superstring} is believed to be UV-finite, and this has been demonstrated up to some order in the early teens in perturbation theory. 
The asymptotic safety program~\cite{weinberg1979ultraviolet} assumes a renormalisable theory of quantum gravity, and derives some of the consequences. The full Lagrangian is not known but some simple examples of Lagrangians illustrate the principle.

Alternatively, instead of quantising gravity as has been achieved for the other fields, we can ask whether spacetime should remain classical.
This is motivated by the fact that the spacetime metric plays a very different role in quantum theory in comparison with other fields. It provides a background causal structure, and notion of time in which the other fields evolve. It is thus reasonable to have lingering questions about whether quantising the metric is appropriate. Only gravity, amongst all the forces, can be described in terms of a universal background geometry. However, for much of the history of this debate, the only theory in which gravity remains classical, was the semi-classical Einstein's equation, where
one replaces the quantum stress-energy tensor of matter (an operator), by its expectation value~\cite{sato1950attempt,moller1962theories,rosenfeld1963quantization}. Treating such an equation as fundamental has long been known to be pathological since it is nonlinear in the density matrix, and leads to super-luminal signalling~\cite{gisin1989stochastic,gisin1990weinberg,polchinski1991weinberg} or a breakdown of the statistical interpretation of the density matrix~\cite{page1981indirect}. It is thus not surprising that equating a classical metric with semi-classical gravity, has led to the rejection of such a possibility~\cite{dewitt1953new,duff1980inconsistency,unruh1984steps,carlip2008quantum}.

However, there are consistent ways to couple classical and quantum systems via a master-equation approach~\cite{blanchard1993interaction,blanchard1995event,diosi1995quantum,poulinKITP,oppenheim2018post,UCLPawula}, and via a measurement and feedback approach in the case of sourcing the Newtonian potential by quantum matter~\cite{kafri2014classical,tilloy2016sourcing,tilloy2017principle}. One can derive the most general form of consistent classical-quantum dynamics, by
demanding that the dynamics preserves the split of classical and quantum degrees of freedom, and preserve the positivity and normalisation of probabilities~\cite{oppenheim2018post,UCLPawula}. This can then be used to construct a consistent master equation for general relativity via the Hamiltonian formulation~\cite{oppenheim2018post,oppenheim2021constraints}. Recently a path integral formulation of classical-quantum dynamics was introduced with Zach Weller-Davies~\cite{oppenheim2023path} and used to formulate a manifestly covariant path integral for classical general relativity coupled to quantum fields~\cite{oppenheim2023covariant}.

Given the existence of a consistent theory of gravity and quantum field theory, we now ask whether the theory is renormalisable.  Renormalisability is especially important for a theory which is motivated by the perspective that gravity is different from the other forces because it describes the geometry of spacetime. If that's the case, we'd expect gravity to be described in terms of the spacetime metric, up to the smallest distance scales. We are therefor most interested in whether the theory of ~\cite{oppenheim2018post,oppenheim2023covariant} is renormalisable in the gravitational degrees of freedom. We will here address this equation by studying the pure gravity theory without coupling to matter. 

We shall find that the path integral formulation of ~\cite{oppenheim2023covariant} is formally renormalisable. In particular, 
 in Section \ref{sec: PIquadratic} 
we will show that the pure gravity
path integral of ~\cite{oppenheim2023covariant} has the same form as that of quadratic gravity~\cite{stelle1977renormalization,stelle1978classical,SalvioQuadratic18,donoghue2021quadratic}. Quadratic gravity is considered an interesting alternative to general relativity, precisely because it is renormalisable~\cite{stelle1977renormalization}. However, the theory suffers from tachyons or ghosts ~\cite{stelle1977renormalization,woodard2023don,SalvioQuadratic18}.  The issue of ghosts is that the Hamiltonian is unbounded from below, leading to the theory being unstable. 
Alternatively, this is equivalent to negative norm states. In fact, for renormalizability, we  \emph{must} trade negative energies for negative norms. There have been attempts to resolve this, this~\cite{mannheim2020ghost,holdom2016quadratic,donoghue2021quadratic, anselmi2017quantum, salvio2018new, salvio2022bicep}, for example by finding that there are meta-stable states in the theory~\cite{Salvio_2019}.
We will see in Section \ref{sec:noghosts} that while the pure gravity theory of~\cite{oppenheim2018post,oppenheim2023covariant} has the same form of action as that of quadratic gravity, it doesn't suffer from tachyons or ghosts. This is precisely because it describes a stochastic classical theory, rather than a unitary quantum theory of gravity. The Hamiltonian for the theory is bounded from below. After explaining this, we next compute the propagator of the theory, which is here interpreted as a correlation kernel. 

While the formal correspondence of the path integral of \cite{oppenheim2023covariant} to that of quadratic gravity, shows that there is a pole-prescription such that the theory is renormalisable, we stop short of claiming that the theory has been shown to be renormalisable, because we have not proven that their is a pole prescription such that the theory is simultaneously renormalisable, and satisfies the property that the dynamics remains completely positive (CP) after renormalisation. Complete Positiviy is a necessary requirement for any dynamics $\mathcal{L}$ of a density matrix, because the density matrix is positive and the dynamics must map positive matrices to positive matrices. Complete positivity is the requirement that $\id\otimes\mathcal{L}$ is a positive map. On non-entangled states such as classical states, CP and Positivity are equivalent.  The other requirement for consistent dynamcis is that the map must preserve the norm, so that probabities are mapped to probabilities. A consequence of the evolution being CP is that correlation functions should be positive semidefinite (PSD) because they are covariance matrices, and variances are positive.  We find strong evidence that the dynamics is CP, by considering the scalar mode, which here corresponds to fluctuations of the Newtonian potential, and finding a pole-prescription, such that the two-point function is positive definite at tree-level. 

In  Appendix \ref{app: mpp}, we also include the matter contribution to the action and give the full solution to the Euler-Lagrange equations, which are different to those of quadratic gravity due to the difference in matter coupling, but also have a very different interpretation. They are not deterministic equations of motion, but rather, are {\it most probable paths}, which as we will shortly explain. They describe extermal stochastic deviations from the equations of motion given boundary conditions. The Appendix also contains an explanation of how the path integral can be gauge fixed in Section~\ref{sec:gauge}, to ensure that the path integral is a genuine sum over geometries. Appendix \ref{sec:twopoint} contains a computation of the acceleration spectral density for different propagators of the theory, which can be used in test the theory in tabletop experiments, or next generation gravitational wave experiments like LISA. These place bounds on the stochastic noise in the gravitational field. 
We conclude in \ref{sec: discussion} with a discussion of experimental constraints and open questions.

\section{The classical-quantum formalism}
\label{sec: CQ}

Let us first briefly review the classical-quantum path integral formalism~\cite{oppenheim2023path} and the path integral for gravity given in~\cite{oppenheim2023covariant}. A classical quantum (CQ) system is represented by a point in phase-space or configuration space $\{q,\dot{q}\}$, and conditional on the classical system being at that point in configuration space, a density matrix living in a Hilbert space  $\mathcal{H}$.  Since we will need to allow for probability distributions over the classical degrees of freedom, a classical-quantum state associates to each classical variable an un-normalized density matrix $\cqstate(q,\dot{q},t) =p(q,\dot{q},t) \hat{\sigma}(\z,t)$ such that $\text{Tr}_{\mathcal{H}}\big(\cqstate(\z)\big) = p(\z,t) \geq 0$ is a normalized probability density over the classical degrees of freedom and $\int dq\, \cqstate(\z,t) $ is a normalized density operator in the Hilbert space. Intuitively, $p(\z,t)$ can be understood as the probability density of being at a point $q,\dot{q}$ and  $\hat{\sigma}(\z,t)$ as the normalized quantum state, given the classical state $\z$ occurs. The dynamics is then consistent, provided it preserves the state space, even when acting on only part of a system, it must map a CQ-state to another CQ-state. This ensures that the classical and quantum degrees of freedom remain classical and quantum respectively. Since the CQ state is positive and normalised, the dynamics must be a completely positive, norm preserving (CPTP) map. The general form of dynamics possessing this property were given in~\cite{oppenheim2018post,UCLPawula} in the master equation picture, and in~\cite{layton2022semi} in the trajectories picture. In the case of path integrals, a sufficient condition for consistent dynamics was given in~\cite{oppenheim2023covariant}, and we here consider a path integral which is of this form. The CQ path integral is related to the Feynman-Vernon path integral for open quantum systems~\cite{FeynmanVernon1963}, and the Onsager-Machlup path integral~\cite{Onsager1953Fluctuations} which appears in the context of stochastic classical dynamics. A table comparing the classical-quantum path integral with these two forms can be found in Appendix~\ref{sec:comparison}.

In order to describe the state of the quantum fields $\phi$ in terms of a density matrix, we formally imagine doubling the number of fields, calling them $\phi^+$ and $\phi^-$ to represent the bra and ket field respectively, as is done in the Schwinger-Keldish or Feynman-Vernon~\cite{FeynmanVernon1963} formalism. We can the write the CQ-state as
\begin{equation}\label{eq: statecomponents}
    \cqstate( \z,t) = \int d \phi^+ d\phi^-\, \cqstate(\z,\phi^+,\phi^-,t) \ | \phi^+ \rangle \langle \phi^- |,
\end{equation}
with $\cqstate( q,\phi^+,\phi^-,t) = \langle \phi^+| \rho(q,t)|\phi^- \rangle $ being the components of the CQ state. In order for the CQ-state to give positive and normalised probabilities of measurement results, $\cqstate(\z,\phi^+,\phi^-,t)$ must be positive semi-definite (PSD) at all points in configuration space, and normalised
\begin{align}
    \int dq\,\text{Tr}(\cqstate( \z,t))=1.
\end{align}

The CQ path integral can then be used to tell us how the components of the density matrix evolve. 
\begin{equation}\label{eq: transition0}
 \rho(q_f,\dot{q}_f,\phi^+,\phi^-,t_f)  = 
   \int   \mathcal{D}q \mathcal{D} \phi^+ \mathcal{D} \phi^- \frac{1}{\mathcal{N}}\,e^{\mathcal{I}_{CQ}[\z,\phi^+,\phi^-,t_i,t_f]}   \rho(q_i,\dot{q}_i,\phi^+_i,\phi^-_i, t_i).
\end{equation}
In Equation~\eqref{eq: transition0} it is implicitly understood that boundary conditions are to be imposed at $t_f$, and we have included a normalization factor $\mathcal{N}$ which can depend on the classical degree of freedom~\cite{UCLnorm2023}. $\mathcal{I}_{CQ}$ is the CQ-action~\cite{oppenheim2023path} which for the purposes of the present work, will be given by
\begin{align}
    \mathcal{I}_{CQ} = i\mathcal{S}_Q[q,\phi^+]-i\mathcal{S}_Q[q,\phi^-]+i\mathcal{S}_{FV}[q,\phi^+,\phi^-] + \mathcal{S}_{diff}[q,\phi^+,\phi^-],
\end{align}
where $i\mathcal{S}_{Q}[q,\phi^\pm]$ are the usual actions for the quantum system which induce unitary evolution on the bra and ket fields $\phi^{\pm}$ respectively. There could be a dependence on the classical degrees of freedom $q$, for example through an interaction term, which describes how the classical system influences the quantum system. $iS_{FV}[q,\phi^+,\phi^-]$ is known as the Feynman-Vernon action. It couples the bra and ket field, which act to decohere the quantum system, and could depend on the classical degrees of freedom, depending on the form of the interaction. A simple example is 
\begin{equation}
    iS_{FV}[\phi^+,\phi^-]=-\frac{D_0}{8}\int dx\,\left((\phi^\rind)^2-(\phi^\lind)^2\right)^2,
\end{equation}
which will cause off-diagonal elements of the density matrix to decay exponentially. The dynamics of the classical degrees of freedom is contained in $\mathcal{S}_{diff}[q,\phi^+,\phi^-]$, which encodes the backreaction of the quantum system on the classical degrees of freedom through a classical-quantum version of the Onsager-Machlup~\cite{Onsager1953Fluctuations} functional. In other words, the classical system diffuses stochastically around its deterministic equations of motion, sourced by the quantum degrees of freedom. For example~\cite{oppenheim2023covariant},

\begin{align}
    \mathcal{S}_{diff}[q,\phi^+,\phi^-]=-\frac{1}{2D_2}\int dx\, \bigg(\Box q-\frac{D_1}{2}\big((\phi^\lind)^2 +(\phi^\rind)^2\big)\bigg)^2,
    \label{eq:boxq}
\end{align}
induces a stochastic force on the classical field $q(x)$, by an amount determined by the average mass density of the bra and ket fields at some point $x$, and whose variance is determined by the constant $D_2$. The constant $D_1$ determines the strength of the back-reaction. Here we might take it to be $D_1=4\pi G_N m$ so that our model is more reminiscent of gravity, for $\q=\Phi$ which behaves as the Newtonian potential. Note that this looks like an equation of motion squared. This action, therefore suppresses paths which deviate too far from the equation of motion $\Box q=\frac{D_1}{2}\left((\phi^\lind)^2 +(\phi^\rind)^2\right)$. This should be compared with the purely classical Onsager-Machlup function
\begin{align}
    \mathcal{I}_{OM}[q]=-\frac{1}{2D_2}\int dt\left(\ddot{q}-\frac{F(q,\dot{q})}{m}\right)^2,
    \label{eq:OM}
\end{align}
which is the path integral for stochastic fluctuations around the force $F$. In the case of more degrees of freedom $q_i$ with equations of motion $m_i\ddot{q_i}-F_i=0$, $D_2^{-1}$ is required to be a positive semi-definite (PSD) matrix. This is because the path integral is computing a probability distribution for initial states to go to final states, or the covariance matrix if we compute the two-point function. Note again, that the Onsager-Machlup (OM) function is essentially an Euler-Lagrange equation squared. The phase-space version of the OM path integral is related to the path integral of Martin-Siggia-Rose (MSR)~\cite{martin1973statistical}.

When $4D_2=D_1D_0^{-1}D_1$ in the CQ path integral, the dynamics can be shown to be completely positive and norm preserving~\cite{oppenheim2023covariant}, and the quantum state remains pure, conditioned on the classical degrees of freedom~\cite{layton2022semi}. This condition saturates an inequality known as the {\it decoherence vs diffusion trade-off}~\cite{oppenheim2022gravitationally}. It ensures that the  cross-terms containing coupling between the $\phi^+$ and $\phi^-$ terms in $S_{FV}$ terms cancels the cross-terms in $\mathcal{S}_{diff}$. Here, these are of the form $(\phi^-)^2(\phi^+)^2$.
One typically also includes source terms in the action, in order to compute correlation functions, but we shall omit these here for the sake of simplicity.

\section{From the classical-quantum path integral to Quadratic Gravity}
\label{sec: PIquadratic}

In the case of classical general relativity coupled to quantum fields, a configuration space path integral was given in~\cite{oppenheim2023covariant} with an action 
 \begin{equation}
 \label{eq:PQG-action}
\begin{split}
    &  \mathcal{I}_{CQ}[ g_{\mu \nu}, \xql,  \xqr] =    \int dx \bigg[   i\big(\mathcal{L}_{Q}[\xqr]  - \mathcal{L}_{Q}[\xql]\big)   -\frac{\text{Det}[-g]}{8}( T^{\mu \nu + } - T^{\mu \nu - }) D_{0,\mu \nu \rho \sigma}(T^{\rho \sigma+ } - T^{\rho \sigma- }) \\
    & - \frac{\text{Det}[-g]}{128 \pi^2 G_N^2}\bigg( G^{\mu \nu} +\Lambda g^{\mu\nu} - 4\pi G_N(T^{ \rho \sigma +} + T^{ \rho \sigma -})\bigg) D_{0, \mu \nu \rho \sigma}\bigg( G^{\rho \sigma} +\Lambda g^{\mu\nu} - 4\pi G_N(T^{ \rho \sigma +} + T^{ \rho \sigma -} \bigg)\bigg],
    \end{split}
\end{equation}
where $\mathcal{L}_Q$ is the standard Lagrangian density for the quantum fields, $T^{\mu\nu\pm}$ are the stress-energy tensor for the $\phi^\pm$ fields, $G^{\mu\nu}$ is the Einstein tensor formed from the metric $g_{\mu\nu}$ in $d$ spacetime dimensions, $g$ is the determinant of the metric, $\Lambda$ is a constant, $G_N$ is Newton's constant and $dx$ is taken to mean integration over $d$ spacetime dimensions. The bottom term is similar to the ''equation of motion squared term'' such as that of Eq.~\eqref{eq:boxq}. If $D_{0, \mu \nu \rho \sigma}$ is positive definite, then paths which violate Einstein's equation sourced by the average stress-energy tensor over the bra and ket fields, are suppressed. The third term is a Feynman-Vernon term which decoheres the field in terms of the stress-energy tensor.

However, there is an apparent tension if one wants to recover all of Einstein's equations in the classical limit. Namely, if we want the action to be manifestly diffeomorphism invariant, then $D_{0, \mu \nu \rho \sigma}$ should be constructed from the metric. 
As noted in~\cite{oppenheim2023covariant}, if $D_{0, \mu \nu \rho \sigma}=D_0 (-g)^{-1/2}g_{\mu\nu}g_{\rho\sigma}$ with $D_0$ a positive constant, geometries which deviate too far from satisfying the trace of Einstein's equations are suppressed. This allows one to construct a diffeomorphism-invariant and self-consistent theory of classical-quantum Nordstrom gravity~\cite{UCLNordstrom} but whether one can recover the remainder of Einstein's equations has been an open question.
The other suggestion in~\cite{oppenheim2023covariant} was for $D_0$ to be given by the generalised deWitt metric parameterised by $\beta$
\begin{align}
\label{eq:dewitt}
D_{0,\mu \nu \rho \sigma}=\frac{D_0}{2\sqrt{-g}}\left(g_{\mu\rho} g_{\nu \sigma}+g_{\mu \sigma} g_{\nu \rho}-2\beta g_{\mu\nu} g_{\rho \sigma}\right).
\end{align} 
This form also consistently recovers the Newtonian limit of general relativity~\cite{layton2023weak}. Unfortunately, in Lorentzian signature, the deWitt metric is not positive semi-definite outside the slow-moving weak field limit of the theory. 

However, as we will see here and in a companion paper~\cite{UCLnorm2023}, the negative eigenvalues of $D_{0, \mu \nu \rho \sigma}$ appear benign.
Namely, we can use the relation 
\begin{align}
R_{\mu \nu} R^{\mu \nu}
=\frac{C^{\mu\nu\rho\sigma}C_{\mu\nu\rho\sigma}}{2}+\frac{R^2}{3}-  
\frac{1}{2}\mathcal{G},
\end{align}
where $C^{\mu\nu\rho\sigma}$ is the Weyl curvature tensor
and $\mathcal{G}$ is the Gauss-Bonnet term, 
to rewrite the pure gravity part of the action $\mathcal{I}^{(gr)}$, as
\begin{align}
 \mathcal{I}^{(gr)}
&= - \frac{1}{128 \pi^2 G_N^2}\int dx \;\text{Det}[-g]\,\left(G^{\mu \nu}+\Lambda g^{\mu\nu}\right)D_{0,\mu \nu\rho \sigma}\left(G^{\rho\sigma} +\Lambda g^{\rho\sigma}\right)\nonumber\\
 &= \int  dx \,\sqrt{-g}\left(-\frac{1}{\R2Term}{R^2}-\frac{ 1}{2\WeylTerm} C^{\mu\nu\rho\sigma}C_{\mu\nu\rho\sigma} +\frac{1}{\Delta_{gb}}\mathcal{G}+\alpha_1 R+\alpha_0)\right)   
 \label{eq:PItoQG}
\end{align} 
with the redefinitions of the coupling constants given by
\begin{align}
 &    \Delta_2 \equiv \frac{128\pi^2G_N^2}{\alpha_2D_0},\quad \alpha_2\equiv\frac{d}{4}-\frac{2}{3}-\beta\left(1-\frac{d}{2}\right)^2, \quad \Delta_{\omega} \equiv\frac{128\pi^2G_N^2}{D_0}, 
     \nonumber\\
&\Delta_{gb}\equiv-\frac{256 \pi^2 G_N^2}{D_0} ,\quad  \alpha_1\equiv\frac{D_0}{128 \pi^2 G_N^2}\Lambda(2-d)(1-\beta d),\quad \alpha_0\equiv\frac{D_0}{128 \pi^2 G_N^2}d\Lambda^2(1-\beta d) \quad.
\label{eq:translation}
\end{align}

Here $\R2Term$,$\WeylTerm$ and $\Delta_{gb}$ are dimensionless, while $\alpha_1$ and $\alpha_0$ have mass dimension $[2]$ and $[4]$ as usual. The Gauss-Bonnet term is a topological invariant in $d=4$ spacetime dimensions and we do not sum over different topologies. Thus, its contribution in the bulk action is a constant and can be ignored. However, as a total divergence it does contribute a boundary term. This is typically discarded, since if the boundary is far enough away, it cannot influence local physics. We shall hold the boundary term fixed here as well, but we want to emphasise that this may change the physical content of the theory, since the final state of the system is not determined by the initial state. Since the Gauss-Bonnet term may be different for different final states, it could influence local physics, no matter how far into the future we place the final boundary. 

Just as in quantum field theory, we can now proceed to divide the action into a free part, which is quadratic in the fields and an interacting part, and then expand the exponent of the interacting part in a power series. The only difference here in comparison to the quantum case, is that we have no $i$ in the action, and we are computing a probability for a transition rather than an amplitude. Just as in quantum field theory, we end up with simple integrals containing powers of fields weighted by an exponent which is quadratic in the fields. The calculation of such integrals leads by Wick's theorem to Feynman rules in coordinate space. In momentum space we encounter the exact same divergent integrals which have to be renormalised. Thus although the metric is classical, physical quantities need to be renormalised because of the theory's stochastic nature -- in the path integral we integrate over classical stochastic fields, and we find the same divergences as if we were computing amplitudes in quantum field theory~\cite{martin1973statistical}.

The theory is power-counting renormalisable (we have included a short summary of power counting renormalisability, in Section \ref{sec:renorm} of the Appendix). Furthermore, up to overall constants and a total divergence, the action is identical to that of quadratic gravity without matter,  which is a renormalisable theory. This suggests that the purely gravitational action of Eq. \eqref{eq:PItoQG} is also renormalisable, since all that has changed is that the coupling constants have an extra factor of $i$, and this will not effect the convergence of integrals. However, there is a caveat here, since the path integral has a different interpretation. Namely it computes a probability distribution, and so, two-point functions of physically measurable quantities need to be positive semidefinite (PSD)~\cite{risken1996fokker,gardiner2004handbook}. This is because they correspond to covariance matrices of probability distributions rather than amplitudes. Renormalisation, and in particular, the pole-prescription, could result in the two-point function not being PSD, or there could be pole-prescriptions which give PSD two-point functions but which break the renormalisability of the theory.  However, we find strong evidence that both conditions are simultaneously satisfiable by finding a pole-prescription which results in a PSD two-point function at tree level, which we discuss in Section \ref{sec:propagator} and Appendix \ref{sec:twopoint}. Since the propagator is also PSD, we expect this property to hold at all orders while retaining the form of the action. 

Let us examine the coupling constants in the theory. Since we want to suppress trajectories which deviate too far from what general relativity predict, we choose $\beta$ so that $\R2Term > 0$ (in $d=4$ this gives $\beta<1/3$).  Further constraints on $\beta$ will follow from the requirement of complete positivity. Note that in quadratic gravity, $\R2Term$ remains positive under RG flow~\cite{julve19781978quantum,
fradkin1981renormalizable,
benedetti2009asymtotic,buccio2024physical}. The Ricci curvature contributions to the Lagrangian density then go as 
$-\left(R-\alpha_1\R2Term\right)^2$ 
which acts to suppress deviations away from some constant Ricci curvature. 

The situation is more complicated for the Weyl curvature term, since it is not of definite sign. However, in~\cite{UCLnorm2023}, we consider the decomposition of the Weyl curvature scalar
\begin{align}
    C^{\mu\nu\rho\sigma}C_{\mu\nu\rho\sigma}=\frac{1}{8}\left(E^\munu E_\munu - B^\munu B_\munu\right),
\end{align}
into its electric $E^\munu$ and magnetic parts $B^\munu$. Both $E^\munu E_\munu$ and $B^\munu B_\munu$ are non-negative, but it's 
only $E^\munu E_\munu$ which contains second time derivatives of the metric. It is only this part of the Weyl curvature contribution which is of the Onsager-Machlup or MSR form, and it acts to suppress paths which have large $E^\munu E_\munu$, as we might expect. On the other hand, the $B^\munu B_\munu$ is non-dynamical. One can choose the discretisation of the path integral into time steps, such that the magnetic part of the Weyl tensor term changes the probability distribution of the initial slice, but because it doesn't contain higher second time derivatives, it is merely part of the normalisation on the initial time slice. This is more easily seen in phase space. By way of analogy, consider the discretised Onsager-Machlup action 
\begin{align}
    I_{OM}=-\delta t \left(\frac{p_k-p_{k-1}}{\delta t}-\frac{1}{m}F(q_{k-1},p_{k-1})\right)^2+\delta t B(p_{k-1},q_{k-1})^2
\end{align}
which acts to evolve an initial distribution at step $k-1$ to one at $k$ a time $\delta t$ later. The normalisation is found by integrating the Gaussian integral over all final states momenta $p_k$, and thus the $B(p_{k-1},q_{k-1})^2$ is normalised away. This forms part of the content of \cite{UCLnorm2023}, and a greater understanding of the implications of this, including whether it effects diffeomorphism invariance, remains outstanding. 

Given its non-dynamical role, it would appear that the positivity of the magnetic Weyl contribution, does not result in paths which deviate far away from Einstein's equation, having too great a weight in the path integral. For the purposes of perturbation theory, we would also appear to be safe. In some sense $\exp\left(B^{\mu\nu}B_{\mu\nu}/16\WeylTerm\right)$, is merely playing the role of a density of states, which can be seen as a modification to Penrose's Weyl curvature conjecture~\cite{penrose1977proc}. The normalisation does not impact renormalisability, because we can first renormalise and then normalise.

The matter part of the action with curvature coupling, is very different to that of quadratic gravity, and is given 
by   
\begin{equation}
\label{eq:matteraction}
\begin{split}
\mathcal{I}^{(m)}=
-\frac{D_0}{16 \pi G_N} \int dx \sqrt{-g} &
\left[-2 \bar{T}^{\mu\nu}R_{\mu\nu}+\left(1+\beta(2-d)\right) \bar{T}R \right.\\
&\left. - 8\pi G_N \beta\bar{T}^2+8\pi G_N \bar{T}^{\mu\nu}\bar{T}_{\mu\nu}+2 (d\beta-1)\Lambda\bar{T}
\right],
\end{split}
\end{equation}
where 
$$
\bar{T}^{\mu\nu}:=\frac{1}{2}\left({T}^{\mu\nu+}+{T}^{\mu\nu-}\right).
$$
This modifies the Euler-Lagrange equations in comparison to those of quadratic gravity, and we include these in Appendix~\ref{app: mpp}. The interpretation of these equations is also very different as we are about to see in the next section. More generally, the action of Eqn \eqref{eq:matteraction} can contain arbitrary coupling constants. The first line of Eq. \eqref{eq:PQG-action} also contains the standard pure quantum action, corresponding to $\pm i\mathcal{L}_Q(\phi^\pm)$ and the decoherence term, which must be considered when considering the full theory with matter.

\section{No negative norm ghosts}
\label{sec:noghosts}
Let us now address the elephant in the room. Although quadratic gravity has garnered widespread interest, owing to it being renormalisable, it suffers from tachyons or negative norm ghosts. This can be traced back to the appearance of second-order derivatives in the action, which was shown by Ostrogradsky to imply that the Hamiltonian is unbounded from below~\cite{ostrogradsky1850memoire}. However, problems arise only under the assumption that the action generates deterministic evolution. Indeed, actions with second-order (time) derivatives often appear when describing classical stochastic processes undergoing Brownian motion. Under the assumption of path-continuity, Brownian motion processes can be described by a stochastic path integral with the probability weighting of each stochastic path given in terms of an action called the Onsager-Machlup function~\cite{Onsager1953Fluctuations}. When describing a process undergoing Brownian motion through kicks in momentum, the configuration space Onsager-Machlup function will include second-order time derivatives when written in terms of the particle position. Let us present a simple example.

Consider the path integral for a free particle undergoing Brownian motion with no drift. The probability of finding a particle at $q(t_f)=q_f$ given that at $t=0$ it was at $q(0)=q_0$ and had velocity $v_0$ and acceleration $a_0$ is given by the Onsager-Machlup path integral
\begin{align}
    P(q_f|q_0,\dot{q}_0,\ddot{q}_0)=\frac{1}{\mathcal{N}}\int_{q_0}^{q_f} \mathcal{D}q\, e^{-\frac{1}{2 D_2}\int_{0}^{t_f} (\ddot{q})^2 dt}.
    \label{eq:PIbrownian}
\end{align}
Note that the path integral acts to suppress the probability of paths which do not satisfy $\ddot{q}=0$, by an amount controlled by the diffusion constant $D_2$. The larger $D_2$ is, the more stochasticity we are likely to find in the paths which are realised.
This is an equivalent description of the dynamics often described by the Langevin equation, $\ddot{q}=F(q)/m+j(t)$, with $F/m$ the drift produced by a deterministic force $F$ (here set to $0$), and $j(t)$ a stochastic white noise process. The dynamics can also be described via the Fokker-Planck equation~\cite{risken1996fokker}  or Ito calculus~\cite{gardiner2004handbook} and we refer the interested reader to~\cite{risken1996fokker} for a derivation of the Onsager-Machlup path integral. The dynamics described by the Onsager-Machlup path integral is in contrast to the Euclidean path integral, often used to compute the partition function. In that case, the Wick-rotation to imaginary time allows one to compute the equilibrium state of a Hamiltonian, while here, we describe diffusion which in the $D_2\rightarrow 0$ limit is deterministic Hamiltonian evolution.

In the case where $-\frac{1}{2D_2}=\frac{i}{\hbar}$, we would be tempted to interpret $\ddot{q}^2$ as a Lagrangian, with Euler-Lagrange equation $d^4 q/dt^4=0$. The path integral of Equation~\eqref{eq:PIbrownian} is then interpreted as generating deterministic unitary evolution of a quantum system. However, 
Lagrangians containing terms second-order or higher in time-derivatives, suffer from the Ostrogradsky instability~\cite{ostrogradsky1850memoire}. Details on the construction of Ostrogradsky and the resulting instability in general are given in detail in~\cite{woodard2015theorem}. For our purposes it suffices to point out that the Hamiltonian which can be constructed for the Lagrangian $\ddot{q}^2$ is $H= P_1X_2+\frac{P_2^2}{4}$, where $P_1=\frac{\partial L}{\partial \dot{q}}-\frac{d}{dt}\frac{\partial L}{\partial \ddot{q}}$, $X_2=\dot{q}$, $P_2=\frac{\partial L}{\partial \ddot{q}}$ and  $X_1=q$. It is linear in $P_1$ and thus unbounded from below, leading to the instability.  
This classical instability must manifest itself quantum mechanically, and indeed it does~\cite{stelle1977renormalization,woodard2015theorem,ganz2021reconsidering}, we either obtain negative energy states (from canonical quantization), or states with negative norms.  

On the other hand, when $D_2$ is positive, the path integral of Equation~\eqref{eq:PIbrownian} corresponds to the free particle Hamiltonian $H=\frac{p^2}{2m}$ with stochastic diffusion around some initial momentum. 
The higher derivative term in the action is no longer a problem, as this is {\it already} a classical probability distribution of a particle with a stable Hamiltonian, and no negative energy-modes. 
The deterministic equation $\ddot{q}=0$, with solution 
 $   q(t_f)=q_0+\dot{q}_0t
 $
is still the global maximum with action $0$, but there are other extremal points of the form $q(t_f)=q_0+\dot{q}_0t+\frac{1}{2}\ddot{q}_0 t^2+\frac{1}{6}\dddot{q}_0 t^3$ which also satisfy $d^4 q/dt^4=0$. However, these are not deterministic paths, but rather {\it the most probable path}~\cite{durr1978onsager,chao2019onsager}, taken for a particle which reaches a $q(t_f)$ and does not necessarily satisfy the deterministic solution $\ddot{q}=0$. 
There is no 
Ostrogradsky instability here, because $d^4 q/dt^4=0$ does not correspond to the deterministic equation of motion generated by a Hamiltonian, but simply tells us the extreme points of the action and thus the most likely trajectories connecting initial and final conditions.

Another way to see this is to see that one can derive the Onsager-Machlup path integral, by integrating out a stochastic noise source which has positive energy (see for example, the derivation in Feynman and Hibbs~\cite{feynman1965quantum} where the $-(\ddot{q})^2$ action is also considered, or the derivation of Yasue~\cite{yasue1978simple} in terms of a Weiner noise process). In contrast, in the quantum theory, one obtains actions of the form $i\ddot{q}^2$, by integrating out a ghost field which must have negative energy (see for example the pedagogical review in \cite{buchbinder2021introduction}). Note that there are additional degrees of freedom of a sort, when one considers a path integral of Onsager-Machlup form as can be seen by the most probably path, which in the case of Brownian motion has additional parameters $\ddot{q}_0$ and $\dddot{q}_0$. This merely represents the fact that the acceleration fluctuates, and thus specifying $q$ and $\dot{q}$ at $t=0$ is not enough to determine the trajectory of the particle. 

This is generic, as we can see from the form of the Onsager-Machlup path integral in Eq. \eqref{eq:OM}. If we vary the Onsager-Machlup action, then since the action is an ''equation of motion squared'', the original equation of motion will be the global extremal point. 
Likewise, for the action of 
Eq.~\eqref{eq:PQG-action}, a solution to the Einstein-Hilbert action is still a global extreme point. The action does not modify the ADM Hamiltonian~\cite{arnowitt2008republication} of general relativity but rather uses the equations of motion generated by it to construct an Onsager-Machlup-like action describing a diffusive process in the metric degrees of freedom.

\section{The propagator and spectral density}
\label{sec:propagator}

The 'propagator' here, is in fact nothing more than the second cumulant of the Onsager-Machlup probability distribution. It also gives  a classical two-point correlation function of the field $h_{\mu \nu}$ where we expand around a flat background: $\sqrt{-g}g_{\mu \nu}=\eta_{\mu \nu}+h_{\mu \nu}$. It has the same form to that of quadratic gravity~\cite{stelle1978classical}  
\begin{equation}
    D^{\delta=0}_{\mu \nu \rho \omega}(p)=
    \frac{1}{(2\pi)^3}\left(\frac{P^{(2)}_{\mu \nu \rho \omega}}{p^2(\Delta_w^{-1} p^2-\alpha_1)}+\frac{P^{(0)}_{\mu \nu \rho \omega}}{p^2[(\frac{1}{2}\Delta_w^{-1}+3\Delta_2^{-1})p^2+\frac{1}{2}\alpha_1]}\right)
%
    \label{eq:twopoint}
\end{equation}
%
%
%
where 
$P^{(n)}$ are the geometric Barnes-Rivers operators~\cite{barnes1963thesis,rivers1964lagrangian,van1973ghost} given in Appendix~\ref{sec:twopoint}
and $\delta$ is a gauge-fixing parameter given in Appendix~\ref{sec:gauge}. Taking the limit $\delta=0$ is equivalent to working in the harmonic gauge $\partial_\nu h^{\mu\nu}=0$. 
We can take the Fourier transform of Eq. \eqref{eq:twopoint} to get the spatial correlation function.  Recall here that the path integral is not computing an amplitude, but rather a probability distribution. Since the two-point function is a covariance kernel in the classical case, it is required to be a positive semi-definite kernel~\cite{risken1996fokker,gardiner2004handbook}. While we only require the dressed two-point function to be PSD, we can ask for the bare propagator to also be PSD, so that the free theory is also well defined. The propagator in position space will be PSD kernel if Eq. \eqref{eq:twopoint} is positive, since Bochner's theorem~\cite{bochner1933monotone} tells us that the Fourier transform of a positive function is PSD. 
Demanding that the propagator be PSD imposes the further condition that not only should $\alpha_2$ in Eq. \eqref{eq:translation} be positive (which in $d=4$ requires $\beta<\frac{1}{3}$), but also that $\alpha_2> 2\alpha_w/3$, and $\alpha_1=0$.
In the quantum case, setting $\alpha_1=0$ would be problematic, since it is equivalent to turning off the Einstein-Hilbert action, but here it merely corresponds to setting $\Lambda=0$ or $\beta=1/d$ as we can see from Equations \eqref{eq:PQG-action} and \eqref{eq:translation}. Since the action should be thought of as ''Einstein's equation squared'' as with Eq \eqref{eq:OM}, we do not expect a term proportional to $R$ unless there is a cosmological constant. 

Indeed with $\alpha_1=0$, one can still recover the Newtonian limit~\cite{layton2023weak} and the Schwarzschild solution. It is likewise also the case that in quadratic gravity one has the Newtonian solutions~\cite{stelle1978classical,alvarez2016aspects} and, in conformal gravity, cosmological solutions~\cite{maldacena2011einstein}.
With $\alpha_1=0$, the action is similar to that of {\it scale-invariant gravity}, which is known to be renormalisable~\cite{salvio2014agravity}, asymptotically free~\cite{julve1978quantum,fradkin1982renormalizable} and, as its name would suggest, scale-invariant and therefore of independent interest~\cite{boulware1983zero,david1984calculability}.
In quantum quadratic gravity, the coefficient $\alpha_w$ is kept negative to protect us from tachyonic ghosts, so that the overall sign of the action is positive.  Here, $\alpha_w$  must be positive for normalisability, but this has no interpretation of a tachyonic particle and is required, in order to get standard diffusive behavior, in actions of the Onsager-Machlup form (Equation \eqref{eq:OM}). The path integral is computing a probability and not an amplitude.

In the quantum theory, the $P^{(2)}_{\mu \nu \rho \omega}$ term corresponds to a spin-two graviton and  $P^{(0)}_{\mu \nu \rho \omega}$ a spin-zero particle, sometimes called the {\it scalaron} when it is quantum and has a mass. In the classical stochastic theory, they respectively correspond to the standard quadrapole gravitational wave, and the scalar stochastic fluctuations of the metric discussed in \cite{layton2023weak,oppenheim2022gravitationally,UCLNordstrom}. One has this additional degrees of freedom, because the Newtonian potential can fluctuate, and so is not completely determined by the matter distribution. The two-point function is different to that used in models where 
the Newtonian potential is sourced by measurements or collapse models~\cite{kafri2014classical,tilloy2016sourcing,tilloy2017principle}. 
In terms of partial fractions the propagator for the spin-2 sector can be written as 
\begin{equation}
    D^{(2)}_{\mu \nu \rho \omega}(p)=\frac{P^{(2)}_{\mu \nu \rho \omega}}{(2\pi)^3}\left(\frac{1}{\alpha_1(p^2-i\epsilon)}+\frac{1}{\alpha_1\Delta_\omega\left(\Delta_\omega^{-1}p^2-\alpha_1-i\epsilon\right)}\right).
    \label{eq:partialfracs}
\end{equation}
with the pole prescription used in the quantum case. In that case, the second term is interpreted as a ghost due to the relative negative sign of the residue. Further, the $-i\epsilon$ prescription, 
gives negative norm states, leading to a loss of unitarity~\cite{stelle1977renormalization,woodard2015theorem}.
We can trade unitarity for renormalizability: the presence of these ghost particles, with the $i \epsilon$ prescription spoils unitarity, and allows for negative norm states. 
In the classical theory, there is no such interpretation. The relative minus sign between the partial fractions, at best, indicates a positive and negative contribution to the correlation kernel, which overall is positive and semi-definite. While the evolution is diffusive, probabilities remain positive and normalised by construction.
Further, the two terms {\it cannot} be interpreted as the propagators of two different particles. Hence, the problems of a theory~\cite{woodard2023don}, with interacting ghosts (particles with negative energies) and non-ghost particles resulting in infinite positive energies, do not arise.

With $\alpha_1$ set to $0$, the propagator is proportional to $1/p^4$ times the Barnes-Rivers operators. The component of the momentum space propagator which goes as $1/p^4$ can be seen to be proportional to the Green's function for the d'Alembertian squared
\begin{align}
    \Box^2 G_{2}(x-x')=\delta^{(4)}(x-x').
    \label{eq:boxsquared}
\end{align}
We want a pole prescription, such that the Green's function is PSD.

In the classical limit of a quantum field theory, the propagator for the transition amplitude  $\langle \phi_f|\phi_i\rangle$ from an initial state to a final state goes as $\frac{1}{p^2+i\epsilon}$, where the $\epsilon$ ensures that positive energies propagate forward in time and negative energies propagate backwards in time.
In the path integral considered here, we are computing a probability, akin to mod-squared $\langle \phi_f|\phi_i\rangle\langle\phi_i|\phi_f\rangle$. Therefore we propose  that the correct $i\epsilon$ prescription is $\frac{1}{p^2+i\epsilon}\frac{1}{p^2-i\epsilon}=\frac{1}{p^4+\epsilon^2}$. We will call this the ''mod-squared'' pole-prescription. The propagator is thus positive in momentum space, and  since the $\epsilon^2$ removes the divergence, Bochner's theorem can be applied and guarantees that the Fourier transform will be a positive definite kernel in position space $x=\vec{x},t$.  The same pole prescription can be applied to the other components of the propagators, which have components of the momentum in the numerators, but only different inverse powers of $p$ in the denominator.

The Fourier transform of $1/p^4$ is performed  in Appendix \ref{sec:twopoint} for various pole prescriptions. We find that the ''mod-squared''
pole prescription described above gives a positive semi-definite two-point function
\begin{equation}
G(x-x') = \frac{D_G}{4\pi^2\epsilon s^2} -\frac{D_G}{32\pi} \sgn(s^2)
\end{equation}
with $s^2=|x-x'|^2-(t-t')^2$, and $D_G$ is the dimensionless coupling constant of the theory, which we relate to the previous conventions in Eq. \eqref{eq:Dg}. Note that outside the lightcone $\Box G(x-x')=0$.
%
%
For completeness, we also compute the two-point function for the Advanced, Retarded, Feynman, and Schwinger-Keldish pole prescription, including for the massive case.  These results are collected in table \ref{tab:Gfuncs}.

We also compute the power spectral density for the acceleration, since this is an observational signature of the theory, measurable in both satellite and tabletop experiments. 
%
%
%
%
%
%
%
We find in \ref{ss:mod} that the acceleration spectral density is
\begin{align}
\label{eq:thespectraldensity}
     S_{aa}(\vec{x},\vec{x}';\omega)|_{\epsilon, \omega\rightarrow 0}=&\frac{D_Gc^3}{4\pi \epsilon |x-x'|}\left(\omega^2 \sin(\frac{\omega}{c}|x-x'|)+\epsilon\cos(\frac{\omega}{c}|x-x'|)\right)
     \nonumber\\
     \approx&\frac{D_Gc^3}{4\pi}\left(\frac{\omega^3}{\epsilon c} +\frac{1}{|x-x'|}\right)
\end{align}
%
where we have put back units of $c$.  
This is computed in the weak field limit $h_{\mu\nu}\approx-2\Phi\delta_{\mu\nu}$ with $\Phi$ the Newtonian potential and such that Eq. \eqref{eq:thespectraldensity} is the spectral density of $\langle\nabla\Phi(x)\nabla\Phi(x')\rangle$. 
Both expressions have an IR divergence as $\epsilon\rightarrow 0$, which is expected since we 
 are
 considering perturbations around Minkowski space, and fluctuations from infinitely far away, and infinitely far in the past will propagate to $x$ and $x'$. If we instead consider perturbations 
 around Friedman-Robertson-Walker, or a universe with a boundary, we do not expect such a divergence. In this case
we expect it to be equivalent to taking  $\epsilon=1/T^2$ with $T$ the light crossing time of the past horizon. This is reasonable since $\sqrt{epsilon}$ acts as a frequency cut-off, and $1/T$ is the minimum possible observable frequency. As we will discuss in Section \ref{sec: exp} taking this to be the Hubble time, appears to be consistent with observation. 

\section{Implications for testing the quantum nature of spacetime }
\label{sec: exp}

Having found the pole-prescription which leads to a PSD two-point correlation function, we can now explore the experimental consequences. We will find that the theory, and by extension the proposition that spacetime is quantum, can be tested in the near term. We do this via the spectral density Eq. \eqref{eq:thespectraldensity}, which encodes the form of stochastic fluctuations in the local Newtonian acceleration. Given any initial condition, if we wait long enough, we expect the correlations of stochastic fluctuations present in the vacuum to have this form. In this way, the classical stochastic field looks very much like a quantum field at short distances, only with a $1/p^4$ two-point function for probabilities instead of amplitudes with a spectrum of $1/p^2$. 

 Fluctuations in acceleration are seen in every precision measurement of force, due to noise in the detector. However, whatever the noise level seen in such a measurement, it will set an upper bound on $D_G$ via Eq. \eqref{eq:thespectraldensity}.
 On the other hand, $D_G$ is lower bounded by decoherence measurements of massive particles in superposition. This can be proven to hold for any theory in which spacetime is classical,  via the ''decoherence vs diffusion trade-off''~\cite{oppenheim2022gravitationally}.  The trade-off between decoherence and diffusion is a requirement that any classical-quantum theory must satisfy, and thus provides a way to rule out such theories, by using both precision gravity measurements, and coherence experiments to squeeze $D_G$ from both sides. So although the difference between quantum and classical fluctuations are subtle, classical fields need to be much larger in comparison to the quantum case.
 
 In this way the theory can be tested, since we do not need to detect fundamental stochastic noise, nor do we need to look for fundamental decoherence, we merely need to put an upper bound on the stochastic noise in the gravitational field, and a lower bound on the coherence time of particles in superposition. If we can do this with sufficient precision, we can rule out the theory.
In comparison to other experimental proposals\cite{bose2017spin,marletto2017gravitationally} this is a more indirect test of the quantum nature of spacetime, since there is some theory input.
Given a particular two-point function for the acceleration, we can construct the corresponding decoherence bound, and there is some freedom in that. However if we demand that the theory is generally covariant, the theory is tightly constrained, since covariance appears to demand that the two-point function depends only on $p^2$, and positivity of the two-point function, would appear to demand that it depend on powers of $p^{4n}$ with $n$ an integer. Hence $1/p^4$ for the scalar fluctuations appears to be relatively model independent, and gives a single dimensionless coupling constant $D_G$ which can be squeezed from both sides. The main caveat is that a mass, induced by coupling the theory to matter would modify the two-point function, as could non-linearities which introduce higher order corrections, but these can presumably be checked.
 If on the other hand, we were to find that the upper and lower bounds can't be pushed past each other, this could be viewed as confirmation of the theory, since there is no reason for a fully quantum theory to satisfy the decoherence-diffusion trade-off.

The precision gravity measurements we will consider~\cite{hoyle2004submillimeter,carney2019tabletop,westphal2021measurement,fuchs2024measuring,janse2024current,armano2018beyond} are typically performed at $\omega \approx 10^{-3} \unit{Hz}$, so the  first term  in Eq \eqref{eq:thespectraldensity}corresponding to $\sin(\omega|x-x'|)/|x-x'|$ is constant over spatial region of order $c/\omega\approx 10^{10}\unit{m}$. While $D_G T^2\omega^3/4\pi$ is significant, the lack of spatial dependence of these fluctuations at the frequencies measured in tabletop experiments mean that this term is not relevant for differential acceleration measurements, which measure differences in the acceleration of two nearby test masses. Additionally these long wavelength fluctuations would appear to effect the entire experiment, and thus we would not expect them to be seen in modern Cavendish experiments. However, we should bear in mind, that accelerations are generally considered to be absolute, with respect to the inertial frame of the system\cite{mach1919science}.

The second term in Eq.~\eqref{eq:thespectraldensity} needs to be integrated over test masses as is done in~\cite{oppenheim2022gravitationally}. For a spherical test mass of radius $R$, this gives $S_{aa}\approx \frac{3D_G}{20\pi R}$.
We can obtain rough order of magnitude bounds from precision acceleration measurements. The experiment of~\cite{westphal2021measurement} uses $\unit{mm}$ 
size masses, and reports $S_{aa}(\omega)\sim 10^{-18}\unit{(m/s^2)^2/Hz}$, leading to 
$D_G\leq 10^{-44}$. 
LISA Pathfinder~\cite{armano2018beyond,armano2024depth} measured $S_{aa}\leq 10^{-30}\unit{(m/s^2)^2/Hz}$ differentially, using $4.5 \unit{cm}$ size test masses, leading to $D_G\leq 10^{-54}$ 
In LISA Pathfinder, the two test masses are $37.6 \unit{cm}$ apart, so we can neglect the correlations in acceleration fluctuations between the test masses. 

We can then use the decoherence vs diffusion trade-off to place a lower bound on $D_G$~\cite{oppenheim2022gravitationally}. A particle in superposition sources the gravitational field, and if spacetime is classical, then the backreaction of the quantum system onto the classical field necessarily causes decoherence~\cite{hall2005interacting,tilloy2016sourcing,poulinKITP,oppenheim2018post}. 
 In the non-relativistic limit,  the decoherence rate $\lambda$ of a particle in superposition of 
two distinct regions of 
volume $V$ is given  by $\lambda=2D_0\frac{M^2}{V}$ with $M$ the mass of the molecule~\cite{oppenheim2022gravitationally}. $V$ here should be understood as the probability density over which the mass of the particle is distributed, in each disjoint region. This is a different rate to that conjectured by Diosi-Penrose, which follows from non-local or non-linear considerations~\cite{diosi1987universal,penrose1998quantum}. Here, the decoherence is due to integrating out the gravitational field~\cite{layton2022semi}. We do not expect relativistic considerations to change this rate.

Coherence measurements such as those of \cite{Gerlich2011} have decoherence rates of $\lambda_{obs}\approx 10\unit{s}^{-1}$, using large organic molecules with total mass $M_{\lambda} = 1.15 \times 10^{-24}kg$ and $N \sim 430$ atoms of size $r\approx 10^{-15}m$. After passing through the slits the molecule becomes delocalized in the transverse direction on the order of $d=2.7\times 10^{-7}m$ before being detected. Since the interference effects are due to the superposition in the transverse direction, which is the direction of alignment of the gratings, we will take the size of the wavepacket of the mass distribution of each atom in the longitudinal direction to be the size of the nucleons $r$. We will also take the verticle direction to be the width of the slit $d\approx 10^{-7} \unit{m}$. 
For macromolecules composed of $N$ particles, and total mass $M$, one finds that the decoherence rate goes as $D_0N^{2/3}M^2/V$ when we take $V \sim 10^{-15} 10^{-7} 10^{-7}m^3 = 10^{-37}m^3$. This can be seen by noting that the nucleons can be considered to be part of a rigid spherical structure -- the total mass $M$ of the molecule when going through the slit, is localised within a volume $V=\frac{d}{N^{1/3}}\times\frac{d}{N^{1/3}}\times r$.
In \cite{oppenheim2022gravitationally}, $D_0N^{1/3}M^2/V$ was used, which is reasonable for long molecules, but if the molecule is spherically packed, then $N^{2/3}$ seems more reasonable.

Since the observed decoherence rate can only be greater than that predicted by the theory, we have
$\lambda_{obs}\geq 2D_0N^{2/3}\frac{M^2}{V}$ which gives  $D_G\approx G_N^2 /D_0 c^3  \geq 10^{-63}$ based on \cite{Gerlich2011}.
The more recent experiment of \cite{fein2019quantum} has $N\approx 2000$, $M\approx 4\times 10^{-23}\unit{kg}$, and an observed coherence time $1/\lambda_{obs}\approx 7.5\times 10^{-3}$ also giving $D_G \geq 10^{-63}$. However, if we imagine the horizontal density is localised in the nucleons $10^{-15}\unit{m}$, then such a localisation would not in principle effect the experiment and would improve the bound to match the upper bound coming from precision gravity measurements. Furthermore, since \cite{Gerlich2011,fein2019quantum} use carbon atoms which are not very dense, and the lower bound on $D_G$ scales like $M^2/V$, going to high density atoms such as gold would enable a much stronger bound by four orders of magnitude. Having large $N$ and large total mass of a macromolecule does not significantly effect the strength of the bound, since it can be compensated for by a decrease in $V$, which could also improve the bound. 

Cesium atoms of mass $10^{-25}\unit{kg}$ have been held in a superposition of different heights over time scales of $60$ seconds and are sensitive to the gravitational potential at different heights~\cite{xu2019probing,panda2024coherence}. Since the longitudinal and transverse directions are irrelevant for these interference experiments, we can imagine that in principle their nucleus is localised to within its radius $\approx 10^{-15}\unit{m}$ in these directions. On the other hand, the atoms are delocalised in height, and so at best we should imagine that the mass of the Cesium atom can be localised to within the diameter of the atom in each arm of the interferometer (we can imagine the atoms could have passed through a grating of size $10^{-10}\unit{m}$ in the horizontal direction). This gives $V=10^{-40}\unit{m^3}$, and similarly, a lower bound on $D_G$ of the same magnitude as the upper bound coming from LISA Pathfinder.

Astrophysical observations may also provide constraints. The first, much larger term $\frac{D_Gc^2}{4\pi}{\omega^3 T^2}$ in the spectral density of Eq. \eqref{eq:thespectraldensity} appears to be consistent with astrophysical observation. Since the variance $\langle(\nabla\Phi)^2\rangle/c^4$ is positive and can be compared to the cosmological constant, we can estimate it's order of magnitude by integrating the spectral density over the relevant frequency band, which is done in Eq. \eqref{eq:IntegratedSpec}. 
\begin{align}
\int^{\omega_2}_{\omega_1} S_{aa}(\vec{x}, \vec{x}'; \omega) \, d\omega \approx \frac{D_G}{4\pi \epsilon |x-x'|^2} \left[ \omega^2 + \frac{2}{|x-x'|^2}  + \epsilon \omega |x-x'| \right]^{\omega_2}_{\omega_1}
\label{eq:IntegratedSpec_main}
\end{align}
If the relevant frequency band is narrow, around a characteristic frequency $\omega_{char}$  the fluctuations accumulate over time $T$ and give a positive density of order $\frac{D_G}{c^2}\omega_{char}^4 T^2$. For the expansion of the universe  the characteristic frequency is $\omega_{char}\approx  1/T$, with $T$ the Hubble time, which would give a negligible effect. For an astrophysical system of mass $M$ and radius $R$ the characteristic frequency is typically taken to be $\omega_{char}\approx \sqrt{\frac{G_NM}{R^3}}$, the free fall time. For galaxies and stars this is typically of order $10^{-16}\unit{s^-1}$ and $10^{-4}\unit{s^-1}$. If the body is effected by fluctuations at a much broader band of frequencies, then these systems may be effected by such fluctuations, and further study of such an effect may be needed. Note that stochastic fluctuations could have accumulated during inflation leading to a positive contribution to the energy density which may act like phantom cold dark matter\cite{oppenheim2024emergence}.






\section{Non-relativistic limit}

In \cite{oppenheim2022gravitationally}, the non-relativistic limit of the local theory considered here was ruled out via the decoherence-vs-diffusion trade-off. We have just seen in Section \ref{sec: exp} that the relativistic theory is consistent with it. The reason is that the finite speed of light means that the volume of stochastic fluctuations which propagate to the points $x$ and $x'$ is one dimension smaller. 

In the non-relativistic limit, Eq \eqref{eq:boxsquared} becomes the biharmonic equation 
\begin{align}
    \nabla^4 G(\vec{x}-\vec{x'})=\delta^{(3)}(\vec{x}-\vec{x'}),
    \label{eq:biharmonic}
\end{align}
with solution~\cite{grunau2010positivity}
\begin{align}
    G(\vec{x}-\vec{x'})=-\frac{1}{8 \pi}|\vec{x}-\vec{x'}|.
    \label{eq:freedom}
\end{align}
This is very different to the two-point functions we have just computed in the relativistic case.
Nonetheless, the non-relativistic acceleration variance, computed from this by applying $\nabla_x\nabla_{x'}$ is given by
\begin{align}
    \nabla_x\cdot\nabla_{x'}G(\vec{x}-\vec{x'})=\frac{1}{4\pi}\frac{1}{|x-x'|}
\end{align}
which indeed corresponds to the finite part of the relativistic acceleration spectral density at low frequency, given in Eq. \eqref{eq:thespectraldensity}.

The Green's function 
\begin{align}
    G_2(\vec{x}-\vec{x'})=\frac{1}{(4\pi)^2}\int d^3y \frac{1}{|\vec{y}-\vec{x}||\vec{y}-\vec{x'}|},
    \label{eq:nofreedom}
\end{align}
was used in~\cite{oppenheim2022gravitationally} to describe correlations in fluctuations of the gravitational field. It can be verified to be a Green's function by direct application of $\nabla^4$. 
Eq. \eqref{eq:nofreedom} is also the non-relativistic limit of the retarded Green's function 
\begin{align}
    G_1(x-y)=-\frac{\delta((t-t')-|x-x'|)\Theta(t-t')}{4\pi |x-x'|},
\end{align}
convoluted with itself.
Both \eqref{eq:freedom} and \eqref{eq:nofreedom} can also be extracted in a formal sense from the work of Diosi-Tilloy~\cite{tilloy2017principle} if a local noise kernel is chosen (in the measurement and feedback approach~\cite{kafri2014classical,tilloy2016sourcing,tilloy2017principle}, the non-local noise kernel $1/4\pi |x-x'|$ is typically used). 

The integral in Eq. \eqref{eq:nofreedom} is in fact infrared divergent (\textit{i.e.} diverges for large $y$). However Eq. \eqref{eq:freedom} can be shown to be equivalent via dimensional regularisation which analytically continues round the divergence. We show this in Appendix \ref{sec:dimreg}. Although \eqref{eq:freedom} by itself is not positive semi-definite, for any finite sized region $|x-x'|$ we can add a large enough constant to it, such that it is PSD. That the resulting two-point function is PSD can be seen via Mercer's theorem, since the kernel is diagonally dominant and with a large enough constant, will always be positive. Since the constant does not affect the variation in acceleration, which is a physical meaningful quantity, this appears to be a gauge choice which we are free to make. 
Furthermore, the variation in the acceleration, \textit{viz.} $\nabla_x \nabla_{x'} G_2(\vec{x}-\vec{x}')$, is an infrared finite (although ultraviolet divergent) integral whose dimensionally continued  result is equal to that computed from the Green's function \eqref{eq:freedom}, as we verify in Appendix \ref{sec:dimreg}.
Although the acceleration two-point function is UV divergent, physically meaningful quantities are finite, since the physically meaningful quantity is the two-point function integrated over test masses\cite{oppenheim2022gravitationally,UCLDMDNE}.

The biharmonic Green's function of Eq. \eqref{eq:freedom} can be shown to be the non-relativistic limit of the propagator computed using the mod-squared pole prescription. Starting from Eq. \eqref{eq:withEpsilon}
\begin{equation}
    G(\vec{x}-\vec{x'},\omega)=\frac{D_G}{4\pi \epsilon|x-x'|}\left(\frac{e^{ik(\omega,\epsilon)|x-x'|}-e^{-ik(\omega,-\epsilon)|x-x'|}}{2i}\right)
    \label{eq:withEpsilon}
\end{equation}
where we have the poles $k(\omega,\pm\epsilon)\equiv\sqrt(\omega^2\pm i\epsilon)\approx \omega\pm i\epsilon/2\omega+\epsilon^2/8\omega^3 + O(\epsilon^3)$. We take the limit $\omega\rightarrow 0$. and then the limit $\epsilon\rightarrow 0$. 

The reason that the relativistic and non-relativistic calculations appear to give such different answers can be understood geometrically. In the Newtonian theory, we have fluctuations $j(y,s)$ at every point in space which can be thought of as sourcing Poisson's equation $\nabla^2\Phi(y)=j(y,s)$. These all instantly cause a change in the local potential at all points $x$ and $x'$, hence there is an IR divergence unless $D_G$ can drop far from the region of interest. This is why the non-relativistic theory could be ruled out in \cite{oppenheim2022gravitationally} via the decoherence-vs-diffusion trade-off when considering a region the size of the Earth. On the other hand, if each fluctuation $j(y,s)$ takes a finite time to reach the points $x$ and $x'$, then it only causes a correlation when the space time interval $s^2(y,x)=(|y-x|^2-(s-t)^2=s^2(y,x')=(|y-x'|^2-(s-t')^2$, which has codimension 1.  Even if the region of interest is the entire universe, the fluctuations in acceleration are still within experimental bounds.

\section{Discussion}
\label{sec: discussion}

Just as the quadratic gravity propagator is here reinterpreted as a classical correlation function,
the saddle points of Equation~\eqref{eq:PItoQG} also have a different interpretation. In the case of quadratic gravity, they are taken to be the Euler-Lagrange equation of motion~\cite{stelle1977renormalization}. Here, as in the Onsager-Machlup case, they correspond to the most probably ''path'' (a spacetime), given an initial and final hypersurfaces. Because the matter part of the action is different to that of quadratic gravity, the most probable paths not only have a different interpretation here, but they also have a different form. In comparison to quadratic gravity, there are various non-minimal couplings between curvature terms and matter degrees of freedom as found in $I^{(m)}$ of Equation \eqref{eq:matteraction}. We compute the most probably paths in the classical limit, by varying the action in Section \ref{app: mpp}.

Let us now summarise the main findings of this work. Together with~\cite{UCLnorm2023}, we have found that although the tensor $D_{0,\mu\nu\rho\sigma}$ is not positive semi-definite, its negative eigenvalues appear benign, since they correspond to a boundary term, or one which can be cancelled by normalisation. The main caveat which should be emphasised, is that we do not fully understand the effect of the  magnetic part of the Weyl curvature by the normalisation, and it could have unintended consequences, for example, the theory may no longer be covariant, although in a manner which is consistent with first picking an initial surface to specify initial data~\cite{UCLnorm2023}. 
Furthermore, while it is standard in the context of higher derivative theories of gravity to drop the Gauss-Bonnet term, since it corresponds to a boundary term, a fuller understanding of the boundary terms in a stochastic theory would be prudent. We have further found that the pure gravity path integral of the classical-quantum theory has the same form as that of quadratic gravity. However, unlike quadratic gravity, the theory doesn't suffer from tachyons or negative norm ghosts. While actions with higher time derivatives terms such as $S_Q=\frac{i}{\hbar}\int dt (\ddot q)^2$ lead to negative norm states in the quantum theory, they are common in classical stochastic theories, and describe diffusive process with Onsagar-Machlup action $S_{OM}=-\frac{1}{2D_2}\int dt (\ddot{q})^2$. The key difference is that the former path integral is computing an amplitude for a deterministic theory with higher derivitive Hamiltonian, while the latter is computing a probability distribution of a standard Hamiltonian with diffusion. In this context one expects the action to be of the form of an equation of motion squared. As such, the $R^2$ and $C^{\mu\nu\rho\sigma}C_{\mu\nu\rho\sigma}$ terms are natural and don't lead to ghosts. We have computed the extremal path corresponding to the action, including coupling to matter. In the stochastic theory, they do not correspond to equations of motion, but rather have the interpretation of most likely spacetimes given an initial and final hypersurface. In the Onsager-Machlup literature, this is often referred to as the ''most probably path''.

Although the action of \cite{oppenheim2023covariant} was constructed in terms of the ''Einstein's equations squared'' we can in retrospect consider the more general form given by Eq~\eqref{eq:PItoQG} without worrying about the translation of coupling constants given by Eqs~\eqref{eq:translation}. Likewise with the matter-curvature couplings of Eq~\eqref{eq:matteraction}. One can then constrain their values by demanding that the two-point function be positive semi-definite and that the path integral be normalisable.  In the classical theory, the propagator of  Eqn~\eqref{eq:twopoint} describes classical correlations in the free gravitational theory which arise due to stochastic fluctuations, rather than an amplitude. Requiring the propagator to be PSD, appears to require setting $\alpha_1$ and $\alpha_0$, the terms corresponding to the bare cosmological constant, to be zero, and singled out the asymptotically free theory. This requirement may not be necessary since the matter couplings are different in this theory in comparison to quadratic gravity 
and we expect further work will clarify whether this condition can be relaxed.

We have further found that the ''mod-squared'' pole prescription, results in the $1/p^4$ propagator being  PSD.  Here, we found an IR divergence which we expect is due to linearising the theory around Minkowski space rather than a universe with a horizon, but more investigation is required. We also  need to 
ensure that both the scalar and tensor mode propagators are PSD. 
A greater understanding of the pole prescriptions in loop diagrams are required, to ensure that the theory is able to be both renormalisable and lead to PSD two-point functions.

The fact that the corresponding quantum theory is asymptotically free is likely to have a number of consequences.
Since the difference between the action of Eq. \eqref{eq:PItoQG}, and that of quadratic gravity, is that the coupling constants are real rather than imaginary, we expect the $\beta$-functions which determine how the coupling constants run are similar to that found in \cite{julve19781978quantum,fradkin1981renormalizable,benedetti2009asymtotic,buccio2024physical}, but with a different sign. This is because the $\beta$-functions depend on the square of the couplings. This suggests that the dimensionless coupling constants $\R2Term$ and $\WeylTerm$ can run to infinity in the UV, so that the diffusion becomes larger at shorter distances. However, even in quadratic gravity, the direction that $\R2Term$ runs depends on the initial value of the coupling constants. This is another area where further work is needed, in order to better understand this behaviour.

If it is confirmed that the diffusion becomes larger at shorter distances than this potentially has implications for the black hole singularity.
While the Schwarszchild solution has zero action and is therefore a global extremum of the action, given the boundary conditions, there are other contributions to the path integral which need to be included~\cite{UCLDMDNE}. This raises the prospect that at short distances, there is no singularity, since the form of the metric become less and less constrained, as the diffusion coefficient gets larger. In effect, the geometry would not need to be Schwarszchild, the closer one got to the singularity. 
It is already believed that singularities in black holes can be avoided or weakened in asymptotically safe gravity (see~\cite{platania2023black} and references therein), and here we conjecture that it is the dominance of fluctuations over the singular deterministic solution of general relativity which is the mechanism behind this. Such a mechanism is not too different from how singularities are removed in quantum electrodynamics.

The modification to gravity at short distances also has implications for tests on the quantum nature of spacetime. These experiments fit into a number of categories. Those based on the detection of gravitationally induced entanglement~\cite{bose2017spin,marletto2017gravitationally} are unaffected by the results presented here since these experiments aim to verify entanglement generation or coherence~\cite{lami2023testing} (c.f.~\cite{kryhin2023distinguishable}) over comparatively long length scales, which would still falsify local classical-quantum theories such as the one considered here, since it cannot generate entanglement. 

A second class of experiments are based on the decoherence-vs-diffusion trade-off\cite{diosi1995quantum,oppenheim2022gravitationally}. As shown in ~\cite{oppenheim2022gravitationally}, any theory in which spacetime is classical, is necessarily constrained in such a way that upper bounds on noise in precision gravity measurements, together with lower bounds on coherence times in interference experiments, could squeeze the theory from both sides. 
The propagators and spectral densities we have computed,  thus provides an experimentally testable signature of the theory. Since precision gravity measurements are able to report the variance in acceleration per frequency (the spectral density), we calculate this quantity in Appendix \ref{sec:twopoint}. While there are caveats due to the possibility of a mass induced from the matter degrees of freedom, Lorentz invariance severely restricts the form of the spectral density. As such we find that the variance in acceleration has a distinctive form as a function of frequency and radius of test mass, and does not depend on its mass, as is typical of many noise processes.  Given current bounds on interference experiments, we have shown that it is feasible that results from LISA Pathfinder, combined with near term coherence experiments involving dense materials, could rule out or confirm the theory.

On the other hand, as emphasised in~\cite{tilloy2017principle, oppenheim2018post,oppenheim2022gravitationally}, experiments based on the decoherence-vs-diffusion trade-off or bounds on anomalous heating can be sensitive to how gravity changes at shorter distances. The non-relativistic models of~\cite{tilloy2016sourcing,tilloy2017principle} are ruled out by both the trade-off~\cite{oppenheim2022gravitationally} and tests of anomalous heating~\cite{tilloy2017principle} unless another length scale is introduced. 
 Indeed, any models such as~\cite{kafri2014classical,tilloy2016sourcing} which are based around sourcing the Newtonian potential or spacetime metric from measurements or stochastic collapse models, are constrained by anomalous heating bounds~\cite{bps,ballentine1991failure,gallis1991comparison,shimony1990desiderata,tilloy2017principle,helou2017lisa,donadi2021underground}. Without a cut-off, which in some models can be as large as the proton radius, such models are ruled out. 
 While the introduction of another length scale may appear artificial, the results of this work suggest that it may be retrospectively prudent. 
 
 In theories such as \cite{oppenheim2018post,oppenheim2023covariant} there is no external collapse model -- the classical nature of spacetime directly induces decoherence in the quantum system, and the dynamics does not reduce to a Lindblad equation if the gravitational degrees of freedom are integrated out. Classicalisation is mediated by spacetime itself. 
Since the gravity sector is related to a quantum theory which is asymptotically free, this supports the view that relativistic models could be well behaved at short distances. 
It is also the case that since $1/\R2Term$ is dimensionless and appears to  run to zero in the UV, one expects to generate another length scale by dimensional transmutation as happens in QCD. This can happen at any scale, and would be determined by experiment. Since gravity is only tested down to the millimeter scale, a better understanding of the short distance behavior of the theory, both experimentally and from a theoretical point of view, could allow for other precision tests of this scale.



Further understanding of the free parameters of the theory could come from CMB data. Since the classical-quantum theory requires an $R^2$ term in the action, and induces an $R$ term once matter degrees of freedom are taken into account, this suggests that the well known model of Starobinsky inflation \cite{starobinsky1980new, vilenkin1985classical,liu2018inflation} could be favoured. Indeed, the Starobinksy model is also favoured by cosmological observations \cite{akrami2020planck}. This would appear to both give further credence to the theory in \cite{oppenheim2018post,oppenheim2023covariant}, and allow for constraints from the CMB on parameters in the theory. Here, one ought to have 
a better understanding of the non-minimal coupling terms in Eq. \eqref{eq:matteraction}. These are expected to induce couplings in $R$~\cite{adler1982einstein} which act as a dynamically generated Planck scale. The aim of this work has been to initiate a better understanding of the pure gravity theory in terms of renormalisability, and relate it to laboratory experiments. While our results suggest that gravity can be described in terms of a classical spacetime to arbitrarily short distances, we have not addressed the question of whether the theory is renormalisable in the matter degrees of freedom. Further investigation is also required, to ensure that the matter couplings do not induce problematic terms in the pure gravitational action.  There are also relationships between the coupling constants, such as the saturation of the decoherence vs diffusion trade-off, which need to be satisfied for the theory to be PSD once quantum matter fields are introduced, and it has to be checked that these can be maintained by the RG flow. This happens is the context of open quantum field theory, although we only have a few examples such as ~\cite{baidya2017renormalization}.

The appearance of terms proportional to the stress-energy tensor contracted with itself in Eq. \eqref{eq:matteraction} are not power-counting renormalisable, unless protected by a symmetry.  Without protection of some kind, the matter sector of the theory may only be an effective theory, valid at low energy in the quantum fields. Fortunately, the non-renormalisable terms may be less relevant at low energy. If we consider a scalar field as an example, the terms which do not appear to be power-counting renormalisable are those which are quadratic in the stress energy tensor. These serve to decohere the quantum field into states which are diagonal in the stress-energy tensor. However, the dominant contribution will come from the mass terms squared, which go as $(\phi^+)^2(\phi^-)^2$ etc. These terms acts to decohere states which have superpositions of different mass density distributions, and they're marginal, appearing with dimensionless coupling constants. On the other hand, the other terms, which have couplings of positive mass dimension, have higher derivatives and would be suppressed at low energy. These act to decoherence states which are in superpositions of very different kinetic energy distributions. In the lab, the dominant back-reaction onto the gravitational field comes from the mass term, rather than the kinetic terms, and so understanding states which have high energy superpositions of different kinetic energy is of far less relevance than understanding the physics of states which are in superpositions of different mass densities.



The motivation to consider classical-quantum theories of gravity is twofold. On the one hand, even in the context of quantum gravity, we are often interested in situations in which we can effectively consider spacetime to be classical. This is the case during inflation, or during black-hole evaporation. There is a sector of the parameter space of classical-quantum dynamics, in which the theory can be taken as an accurate description of a fully quantum theory in which one subsystem (here spacetime), has decohered\cite{UCLQQtoCQ}. Whether that parameter space is a close enough approximation of the theory considered here, is an open question. However, these renormalisability results could motivate us to consider such a limit, or the use of path-integral methods to compute probabilities rather than amplitudes. Whether such a computation can be made renormalisable is an interesting open question.

{\bf Acknowledgements}
JO would like to thank Daine Danielson, Lajos Diosi, Sigbjørn Hervik, Alex Maloney, John Moffat, Andrew Pontzen, Gautam Satishchandran, Bob Holdom, Andy Svesko, Antoine Tilloy, and Zach Weller-Davies for helpful discussions, and the Perimeter Institute for hosting him while part of this work was being conducted. 
We also thank Alberto Salvio, Emanuele Panella and Isaac Layton for valuable discussion.
 JO was supported by the Simons Foundation {\it It from Qubit} Network. A.R acknowledges financial support from UKRI.
AG and MS would like to thank University College London for hosting them while part of this work was being conducted.

\bibliography{refcq,refRenorm}

\newpage
\appendix
\section{Comparison of classical, quantum, and classical-quantum path integrals}
\label{sec:comparison}

The classical-quantum path integral generalizes the Feynman-Vernon path integral of open quantum systems and the stochastic path integral of classical systems. In ~\cite{oppenheim2023path}, we compare and contrast these three different path integrals, and for convenience, we include Table~\ref{tab: pathintegrals} below.\begin{table}[H]

\caption{\label{tab: pathintegrals}A table representing the classical, quantum, and classical-quantum path integrals.}
\small
\begin{subtable}{\linewidth}
\begin{tblr}{
  colspec = {X[2cm,c]X[c]},
  stretch = 0,
  rowsep = 4pt,
  hlines = {black, 0.5pt},
  vlines = {black, 0.5pt},
}
 & Classical stochastic  \\

    \parbox{\linewidth}{Path integral} & $p(q, p,t_f) = \int \mathcal{D} q \mathcal{D} p \ e^{iS_C[q,p]} \delta( \dot{q} - \frac{\partial H}{\partial p}) p(q,p,t_i) $  \\ 

\parbox{\linewidth}{Action} & $iS_C =-\int_{t_i}^{t_f} dt \frac{1}{2} \ (\frac{\partial H}{\partial q} + \dot{p}) D_2^{-1} (\frac{\partial H}{\partial q}  + \dot{p}) $  \\ 

\parbox{\linewidth}{CP condition} &     $D^{-1}_2$ a positive (semi-definite) matrix, $D^{-1}_2 \succeq 0 $  \\

\end{tblr}

\caption{The path integral for continuous, stochastic phase space classical dynamics ~\cite{Onsager1953Fluctuations, freidlin1998random,Weber_2017, Kleinert}. One sums over all classical configurations $(q,p)$ with a weighting according to the difference between the classical path and its expected force $-\frac{\partial H}{\partial q}$, by an amount characterized by the diffusion matrix $D_{2}$. In the case where the force is determined by a Lagrangian $L_c$, the action $S_C$ describes suppression of paths away from the Euler-Lagrange equations $iS_C =-\int_{t_i}^{t_f} dt \frac{1}{2} (\frac{\delta L_c }{\delta q_i}) (D_2^{-1})^{ij} (\frac{\delta L_c }{\delta q_j}) $, by an amount determined by the diffusion coefficient $D_2$. The most general form of classical path integral can be found in ~\cite{Weber_2017, Kleinert, oppenheim2023path}  } 
\end{subtable}
\vspace{0.1cm}

\begin{subtable}{\linewidth}
\begin{tblr}{
  colspec = {X[2cm,c]X[c]},
  stretch = 0,
  rowsep = 4pt,
  hlines = {black, 0.5pt},
  vlines = {black, 0.5pt},
}
 & Quantum    \\

    \parbox{\linewidth}{Path integral} & $\rho(\phi^{\pm}, t_f) = \int \mathcal{D} \phi^{\pm} \ e^{iS[\phi^+] - iS[\phi^- ] + iS_{FV}[\phi^+, \phi^-] } \rho(\phi^{\pm}, t_i)$   \\ 

\parbox{\linewidth}{Action} & $\displaystyle 
\begin{aligned}
& S[\xq]  = \int_{t_i}^{t_f} dt \big( \frac{1}{2}\dot{\xq}^2 + V(\xq) \big), \ \ \ iS_{FV} =  \int_{t_i}^{t_f} dt \big ( D^{\alpha \beta}_0  \lin_{\alpha}^{\rind} \lin_{\beta}^{* \lind} - \frac{1}{2} D^{\alpha \beta}_0 (L^{* \lind}_{\beta}\lin_{\alpha}^{\lind} + L^{*\rind}_{\beta}\lin_{\alpha}^{\rind} ) \big) 
 \end{aligned}$ \\

\parbox{\linewidth}{CP condition} &   $D_0^{\alpha \beta}$ a positive (semi-definite) matrix, $D_0 \succeq 0$. \\

\end{tblr}

\caption{\label{tab: quantumPI} A path integral for a general autonomous quantum system, here taken to be $\phi$. The quantum path integral is doubled since it includes a path integral over both the bra and ket components of the density matrix, here represented using the $\pm$ notation. In the absence of the Feynman Vernon term $S_{FV}$ ~\cite{FeynmanVernon1963}, the path integral represents a quantum system evolving unitarily with an action $S[\phi]$. When the Feynman Vernon action $S_{FV}$ is included, the path integral describes the path integral for dynamics undergoing Lindbladian evolution ~\cite{Lindblad:1975ef, GKS} with Lindblad operators $L_\alpha(\phi)$. Because of the $\pm$ cross terms, the path integral no longer preserves the purity of the quantum state, and there will generally be decoherence by an amount determined by $D_0$. As an example, taking  $\lin^{\pm}=\phi^\pm (x)$ a local field, and  
 $D_0^{\alpha \beta}=D_0$ results in a Feynman-Vernon term
$iS_{FV} =  - \frac{1}{2}D_0 \int_{t_i}^{t_f} dt dx \big (\phi^\lind(x)-\phi^\rind(x)\big) ^2$ which decoheres the state in the $\phi(x)$ basis, since off-diagonal terms in the density matrix, where $\phi^\rind(x)$ is different to $\phi^\lind(x)$, are suppressed.
}
\end{subtable}

\vspace{0.3cm}

\begin{subtable}{\linewidth}
\begin{tblr}{
  colspec = {X[2cm,c]X[c]},
  stretch = 0,
  rowsep = 4pt,
  hlines = {black, 0.5pt},
  vlines = {black, 0.5pt},
}
 & Classical-quantum   \\

    \parbox{\linewidth}{Path integral} & $ \rho(q,p, \phi^{\pm}, t_f) = \int \mathcal{D} q  \mathcal{D} p \mathcal{D} \phi^{\pm} \ e^{iS_C[q,p] + iS[\phi^+] - iS[\phi^{-} ] + iS_{FV}[\phi^{\pm}]+ iS_{CQ}[q,p, \phi^{\pm}]  } \delta( \dot{q} - \frac{p}{m}) \rho(q,p,\phi^{\pm}, t_i)
    $\\ 

\parbox{\linewidth}{Action} & $\displaystyle 
\begin{aligned}
& iS_C[z] + iS_{CQ}[z,\phi^{\pm}] = -\frac{1}{2} \int_{t_i}^{t_f} dt \ D_2^{-1}\big( \frac{\partial H_c}{\partial q}+ \frac{1}{2} \frac{\partial V_I[q,\phi^+]}{\partial q } + \frac{1}{2} \frac{\partial V_I[q,\phi^-]}{\partial q } + \dot{p}\big)^2.
 \end{aligned}$ \\

\parbox{\linewidth}{CP condition} &   $D_0 \succeq 0, D_2 \succeq 0$ and $4D_2 \succeq  D_0^{-1}$ \\

\end{tblr}
\caption{A phase space path integral for continuous, autonomous classical-quantum dynamics. The path integral is a sum over all classical paths of the variables $\z$, as well as a sum over the doubled quantum degrees of freedom $\phi^{\pm}$. The action contains the purely quantum term from the quantum path integral in Table~\ref{tab: quantumPI}, but also includes the term $iS_C+iS_{CQ}$. This suppresses paths away from the averaged drift, which is sourced by both purely classical terms described by the Hamiltonian $H_c$ and the back-reaction of the quantum systems on the classical ones, described by a classical-quantum interaction potential $V_I$. The most general form of classical-quantum path integral can be found in ~\cite{oppenheim2023path}. In order for the dynamics to be completely positive, the decoherence-diffusion trade-off $4D_2 \succeq D_0^{-1}$ must be satisfied ~\cite{diosi1995quantum,UCLPawula,oppenheim2022gravitationally}, where $D_0^{-1}$ is the generalized inverse of $D_0$, which must be positive semi-definite. When the trade-off is saturated, the path integral preserves the purity of the quantum state, conditioned on the classical degree of freedom ~\cite{layton2022semi}.
}
\end{subtable}
\label{tab: MasterEquationTable} 
\end{table}

\section{Propagators and spectral densities}

\label{sec:twopoint}

We consider the propagators of Equation \eqref{eq:twopoint}, and first consider the massless case, since this appears to be required for both the scalar and tensor mode to have a PSD two-point function. However, matter coupling could add a small mass while still preserving the PSD property. We will here compute the two-point function for the scalar mode in position space, and the power spectral density of the acceleration, which can be used in tabletop experiments which measure variations in acceleration. We will consider both the $\frac{1}{p^4+\epsilon^2}$ prescription, and the advanced and retarded pole prescriptions.. Table \ref{tab:Gfuncs} collects these results. The similarity to the Schwinger-Keldysh prescription is discussed in \eqref{ss:SK}.

\begin{table}[H]
\caption{\label{tab:Gfuncs}
The Fourier transform of the two-point function $\frac{1}{p^4}$ for different pole-prescriptions, and the corresponding acceleration spectral density.
The massless Feynman propagator is $\frac{1}{4\pi^2is^2}$ where  $s^2 \equiv (\vec{x}-\vec{x}')^2-(t-t')^2$}
\small
\begin{subtable}{\linewidth}
\begin{tblr}{
  colspec = {X[2cm,c]X[c]},
  stretch = 0,
  rowsep = 4pt,
  hlines = {black, 0.5pt},
  vlines = {black, 0.5pt},
}
Type & Green's functions & Spectral Density \\

\parbox{\linewidth}{Retarded} & 
$ G_R(x-x') = -\frac{1}{8\pi}\theta(t-t'-|\vec{x}-\vec{x}'|) $ & 
$ S_{aa,R}(\vec{x},\vec{x}',\omega) = \frac{D_G}{ 4\pi } \left(\frac{1}{|x-x'|}+\frac{i \omega}{2}\right) e^{i \omega |x-x'|} $  \\ 

\parbox{\linewidth}{Advanced} & 
$ G_A(x-x') = \frac{1}{8\pi}\theta(t-t'+|\vec{x}-\vec{x}'|) $ & 
$ S_{aa,A}(\vec{x},\vec{x}',\omega) = \frac{D_G}{ 4\pi } \left(\frac{1}{|x-x'|}-\frac{i \omega}{2}\right) e^{-i \omega |x-x'|} $ \\

\parbox{\linewidth}{Mod-squared prescription} & 
$ G(x-x') = \frac1{4\pi^2\epsilon s^2} - \frac{1}{32\pi} \sgn(s^2)\ $ &  
$ S_{aa}(\vec{x},\vec{x}';\omega)=
\frac{D_G}{4\pi \epsilon |x-x'|}\left(\omega^2 \sin(\omega|x-x'|)+\epsilon\cos(\omega|x-x'|)\right)$

\end{tblr}
\end{subtable}
\end{table}

Starting from Eq. \eqref{eq:twopoint}, we
take $\alpha_1=0$ so that both the scalar and tensor fluctuations are positive in momentum space give
\begin{equation}
    D^{\delta=0}_{\mu \nu \rho \omega}(p)=
    \frac{1}{(2\pi)^3 p^4}\left(\frac{P^{(2)}_{\mu \nu \rho \omega}}{\Delta_w^{-1}}+\frac{P^{(0)}_{\mu \nu \rho \omega}}{\frac{1}{2}\Delta_w^{-1}+3\Delta_2^{-1}}\right)
\label{eq:twopoint1}
\end{equation}
where the (Barnes-Rivers) operators are given by~\cite{stelle1977renormalization}
\begin{equation}
    P^{(2)}_{\mu \nu \rho \omega}=\frac{1}{2}(\theta_{\mu\rho}\theta_{\nu\omega}+\theta_{\mu\omega}\theta_{\nu\rho})-\frac{1}{3}\theta_{\mu \nu}\theta_{\rho \omega},
    \label{Proj-2}
\end{equation}
\begin{equation}
\label{Proj-0}
    P^{(0)}_{\mu \nu \rho \omega}=\frac{1}{3}\theta_{\mu \nu}\theta_{\rho \omega},
\end{equation}
\begin{equation}
    \theta_{\mu\nu}=\eta_{\mu \nu}-\frac{p_{\mu}p_{\nu}}{p^2}
\end{equation}
and $p$ is the 4-momentum, and $\eta$ the Minkowski metric. 

In the following analysis we will only consider the  momentum-free part of these projectors as this is relevant for current tabletop experiments. 
\begin{align}
    P^{(0)}_{\mu \nu \rho \omega,sym}=\frac{1}{3}\left(\frac{11}{12}\eta_{\mu \nu}\eta_{ \rho \omega} +\frac{1}{24}\eta_{\mu \rho}\eta_{\nu \omega}+\frac{1}{24}\eta_{\mu \omega}\eta_{\nu \rho}\right)
\end{align}
\begin{align}
      P^{(2)}_{\mu \nu \rho \omega,sym}= \frac{3}{2}(P^{(0)}_{\mu \rho \nu \omega,sym}+ P^{(0)}_{\mu \omega \nu \rho,sym})-P^{(0)}_{\mu \nu \rho \omega,sym}
\end{align}

Then the Fourier transform of the momentum dependence we denote by 
\begin{align}
    G^{(4)}(x,x')=
    \frac{1}{(2\pi)^3}\left(\frac{P^{(2)}_{\mu \nu \rho \omega}}{\Delta_w^{-1}}+\frac{P^{(0)}_{\mu \nu \rho \omega}}{\frac{1}{2}\Delta_w^{-1}+3\Delta_2^{-1}}\right)\frac{1}{ (2\pi)^4 }\int d^3kd\omega \frac{ e^{i\vec{k}\cdot(\vec{x}-\vec{x}')-i\omega (t-t')}}{\left((\vec{k},\omega)\cdot (\vec{k},\omega)^T\right)^2 }
\label{eq:Fourier1}
\end{align} 
where we write the momentum $4$-vector $p$ in terms of $\vec{k},\omega$. Henceforth, we will suppress the constant factor upfront.
The propagator is the Green's function for $\Box^2$. 



We will first perform the Fourier transform over $\vec{k}$ only, since from the point of view of experiments, we are interested in computing the spectral density as a function of frequency $\omega$. I.e. we first compute
\begin{align}
   G(\vec{x},\vec{x}',\omega)=&\frac{1}{ (2\pi)^3}\int\frac{d^3k e^{i\vec{k}\cdot(\vec{x}-\vec{x}')}}{|\vec{k}|^4-2|\vec{k}|^2\omega^2+\omega^4}
\label{eq:phispec}
\end{align}

Writing $d^3k=k^2\sin{\theta}d\theta d\phi dk$ and $\vec{k}\cdot (\vec{x}-\vec{x})=k|\vec{x}-\vec{x}'|\cos\theta$, $k\equiv |\vec{k}|$, and using 
\begin{align}
\int_0^\pi d\theta \sin\theta e^{ikx\cos\theta}=\frac{2\sin(kx)}{kx}
    \label{eq:int1}
\end{align}
gives
\begin{align}
   G(\vec{x},\vec{x}',\omega)=&\frac{2}{ (2\pi)^2 |\vec{x}-\vec{x}'|}\int_0^\infty dk\frac{ k\sin{k|x-x'|}}{|\vec{k}|^4-2|\vec{k}|^2\omega^2+\omega^4}
\label{eq:int3}
\end{align}


Since the physically meaningful quantity in terms of tabletop experiments is the acceleration two-point function, we can take spatial derivatives of Eq \eqref{eq:phispec}, to bring down another $k^2$. This would give
the variance in acceleration as a function of $\omega$, which is the spectral density $S_{aa}(\vec{x},\vec{x}';\omega)$. We are therefore also interested in the quantity
\begin{align}
  S_{aa}(\vec{x},\vec{x}';\omega):=&D_G\sum_i \frac{\partial^2 G(\vec{x},\vec{x}',\omega)}{\partial x_i\partial x'_i}\nonumber\\
 \label{eq:spectdensity}
\end{align}
where we include a constant of proportionality $D_G$ which we will determine later, so that the spectral density is computing the power spectral density of the variance in the Newtonian acceleration. Substituting Eq. \eqref{eq:phispec}, into Eq. \eqref{eq:spectdensity} we obtain
\begin{align}
  S_{aa}(\vec{x},\vec{x}';\omega):=&\frac{2D_G}{ (2\pi)^2 |x-x'|}\int_0^\infty\frac{dk |\vec{k}|^3\sin{k|x-x'|}}{|\vec{k}|^4-2|\vec{k}|^2\omega^2+\omega^4}
  \label{eq:spectralAtx}
\end{align}

We now consider several pole-prescriptions.

\subsection{Mod-squared ($+\epsilon^2$) pole prescription}
\label{ss:mod}
We first consider the following pole prescription for the propagator
\begin{equation}
    \label{eq: plus_minus_prescription}
    G(\vec{k},\omega)=\frac{1}{(\vec{k^2}-\omega^2-i\epsilon)(\vec{k^2}-\omega^2+i\epsilon)}=\frac{1}{2i\epsilon}(\frac{1}{\vec{k^2}-\omega^2-i\epsilon}-\frac{1}{\vec{k^2}-\omega^2+i\epsilon})
\end{equation}
which we justified in Section \ref{sec:propagator} in terms of computing probabilities. 

Taking the fourier transform, 
\begin{align} G(\vec{x}-\vec{x'},\omega)=
    \frac{1}{ (2\pi)^3(2i\epsilon) }\int d^3ke^{i\vec{k}\cdot(\vec{x}-\vec{x}')-i\omega (t-t')}(\frac{1}{\vec{k^2}-\omega^2-i\epsilon}-\frac{1}{\vec{k^2}-\omega^2+i\epsilon})
\end{align}
We can shift to spherical coordinates, and then integrate over the solid angle to obtain,
\begin{equation}
   G(\vec{x}-\vec{x'},\omega)=\frac{2}{(2i\epsilon)(2\pi)^2|x-x'|}\int_0^{\infty} k dk \sin\left({k|x-x'|}\right)(\frac{1}{\vec{k^2}-\omega^2-i\epsilon}-\frac{1}{\vec{k^2}-\omega^2+i\epsilon})
\end{equation}

We close the contour in the upper half plane, where we have the poles $k(\omega,+\epsilon)\equiv\sqrt{\omega^2+i\epsilon}\approx \omega+i\epsilon/2\omega+\epsilon^2/8\omega^3 + O(\epsilon^3)$ for the first partial fraction, and $-k(\omega,-\epsilon)\equiv-\sqrt{\omega^2-i\epsilon}\approx-\left( \omega-i\epsilon/2\omega+\epsilon^2/8\omega^3 + O(\epsilon^3)\right)$ for the second. We thus obtain
for Eq. \eqref{eq:int3}
\begin{equation}
    G(\vec{x}-\vec{x'},\omega)=\frac{1}{4\pi \epsilon|x-x'|}\left(\frac{e^{ik(\omega,\epsilon)|x-x'|}-e^{-ik(\omega,-\epsilon)|x-x'|}}{2i}\right)
    \label{eq:withEpsilon}
\end{equation}

Taking the Fourier transform of \eqref{eq:withEpsilon}, the analysis depends on whether one chooses time-like or space-like arguments. If we choose time-like arguments we can boost to a frame in which $\vec{x}-\vec{x}'\to0$. Then the Fourier transform is 
\begin{align}
    G(0,t-t') = \frac1{8\pi\epsilon}\int^\infty_{-\infty}\frac{d\omega}{2\pi}\left[ k(\epsilon,\omega)+k(-\epsilon,\omega)\right] e^{-i\omega (t-t')}\,.
\end{align}
Deforming the contour of integration to go around the cut in $k(\epsilon,\omega)$, the integral of the first term in brackets can be evaluated in terms of the $K_1$ modified Bessel function of the second kind. Evaluating the second term in a similar way, we thus find
\begin{align}
\label{Gzerot}
    G(0,t-t') = -\frac{i}{8\pi^2|t-t'|\sqrt{\epsilon}} \left\{ e^{-i\pi/4}K_1\left(e^{i\pi/4}\sqrt{\epsilon}|t-t'|\right) - e^{i\pi/4}K_1\left(e^{-i\pi/4}\sqrt{\epsilon}|t-t'|\right)\right\}\,.
\end{align}
Expanding in small $\epsilon$, using
\begin{align}
\label{KoneSeries}
    K_1(z) = \frac1z +\left(\frac{\gamma_E}{2}-\frac14+\frac12\ln\frac{z}2\right)z + O(z^3\ln z)\,,
\end{align}
and boosting back we get, for time-like separated space-time points $(\vec{x},t)$, $(\vec{x}',t')$ and up to terms vanishing in $\epsilon$,
\begin{align}
\label{Gfinal}
    G(\vec{x}-\vec{x'},t-t') = \frac1{4\pi^2\epsilon s^2}+\frac1{32\pi}\,,
\end{align}
where we have expressed it in terms of the space-time interval:
\begin{align}
    s^2 = -(t-t')^2+(\vec{x}-\vec{x}')^2\,.
\end{align}
On the other hand, if we have space-like arguments ($s^2<0$), we can boost to a frame in which $t-t'=0$. Then the Fourier transform is given directly by the integral over $\omega$ of   \eqref{eq:withEpsilon} (multiplied by $1/2\pi$). For the first exponential in \eqref{eq:withEpsilon}, the integral becomes absolutely convergent if we rotate the integration contour by $\pi/4$. This results in another integral representation of $K_1$. Evaluating the integral over the second exponential in similar fashion, the result is just minus \eqref{Gzerot} with $t-t'$ replaced by $|\vec{x}-\vec{x}'|$. Thus, expanding in $\epsilon$ and boosting back, we get the first term in \eqref{Gfinal} but now for space-like arguments and $-1/32\pi$ for the constant term. We conclude that, up to terms that vanish in $\epsilon$ and for $s^2\ne0$,
\begin{align}
    G(\vec{x}-\vec{x'},t-t') = \frac1{4\pi^2\epsilon s^2} -\frac1{32\pi}\sgn(s^2)\,.
\end{align}

The spectral density can be computed by performing the integral in  Eq. \eqref{eq:spectralAtx}
\begin{equation}
     S_{aa}(\vec{x},\vec{x}';\omega)=\frac{D_G}{4\pi \epsilon |x-x'|(2i)}\left(k^2(\omega,+\epsilon)e^{ik(\omega,+\epsilon)|x-x'|}-k^2(\omega,-\epsilon)e^{-ik(\omega,-\epsilon)|x-x'|}\right)
     \label{eq:NewSpec}
\end{equation}

We can find the non-relativistic limit of the spectral density, by Taylor expanding \eqref{eq:NewSpec} in powers of $\epsilon$.
With $\epsilon$ being small, but finite, we have:

\begin{align}
    S_{aa}(\vec{x},\vec{x}';\omega)=&\frac{D_G}{4\pi \epsilon |x-x'|}\left(\omega^2(\sin\omega|x-x'|)(1- \frac{\epsilon|x-x'|}{2\omega}+\frac{\epsilon^2|x-x'|^2}{8\omega^2})+
   \epsilon \cos(\omega|x-x'|)(1-\frac{3\epsilon|x-x'|}{8\omega})\right)
    \nonumber
\end{align}

We now take $\epsilon$  of order $\omega^3$, and, discarding terms of combined order $O>0$, 
we obtain

\begin{align}
\label{eq:nonrel}
     S_{aa}(\vec{x},\vec{x}';\omega)|_{\epsilon, \omega\rightarrow 0}=&\frac{D_G}{4\pi \epsilon |x-x'|}\left(\omega^2 \sin(\omega|x-x'|)+\epsilon\cos(\omega|x-x'|)\right)
     \nonumber\\
     \approx&\frac{D_G}{4\pi}\left(\frac{1}{|x-x'|}+\frac{\omega^3}{\epsilon}\right)
\end{align}


We can integrate this over some frequency bandwidth, provided the lower frequency $\omega_1$ satisfies $\omega_1^2\gg \epsilon$
\begin{align}
\int^{\omega_2}_{\omega_1} S_{aa}(\vec{x}, \vec{x}'; \omega) \, d\omega &= \frac{D_G}{4\pi \epsilon |x-x'|^2} \left[ -\omega^2 \cos(\omega |x-x'|) + \frac{2\omega}{|x-x'|} \sin(\omega |x-x'|) + \frac{2}{|x-x'|^2} \cos(\omega |x-x'|) + \epsilon \sin(\omega |x-x'|) \right]^{\omega_2}_{\omega_1}\nonumber\\
&\approx \frac{D_G}{4\pi \epsilon |x-x'|^2} \left[ \omega^2 + \frac{2}{|x-x'|^2}  + \epsilon \omega |x-x'| \right]^{\omega_2}_{\omega_1}
\label{eq:IntegratedSpec}
\end{align}

When computing the two-point function, the propagator will be multiplied by the dimensionless coupling constant $D_G$ which we now compute such that this gives the two-point function for the Newtonian potential $\Phi$ (see for example \cite{layton2023weak,UCLDMDNE}). We now relate $D_G$ to the other coupling constants used in this paper. Recall Eq. \eqref{eq:twopoint} i.e.
\begin{equation}
\langle h_{\mu\nu}h_{\rho\omega}\rangle-\langle h_{\mu\nu}\rangle\langle h_{\rho\omega}\rangle=
    \frac{1}{(2\pi)^3}\left(\frac{P^{(2)}_{\mu \nu \rho \omega}}{\Delta_w^{-1}}+\frac{P^{(0)}_{\mu \nu \rho \omega}}{\frac{1}{2}\Delta_w^{-1}+3\Delta_2^{-1}}\right)\frac{1}{p^4}
    \label{eq:hmunuTwoPoint}
\end{equation}
and contract the indices of the Barnes-River operators 
pairwise,  $P^{\mu\rho}_{\mu\rho}=\frac{1}{3}\theta^\mu_\mu \theta^\rho_\rho=3$, where each $\theta^\mu_\mu=3$. 
Focusing on the scalar propagator, we substitute $d=4$ into $\alpha_2$  to get 
\begin{equation}
     \Delta_2 = \frac{128\pi^2G_N^2}{\left(\frac{1}{3}-\beta\right)D_0}, \quad \Delta_{\omega} =\frac{128\pi^2G_N^2}{D_0}.
\end{equation}
when substituting the $\Delta$s back in the prefactor propagator, it becomes 
\begin{equation}
    \frac{1}{(2\pi)^3}\frac{P^{(0)}_{\mu \nu \rho \omega}}{\frac{1}{2}\Delta_w^{-1}+3\Delta_2^{-1}} = \frac{32 G^2_N P_{0,\mu\nu\rho\sigma}}{3 D_0  \pi (1 - 2 \beta)}.
\end{equation}
At this point, contracting the indices of the projector from Eq.~\eqref{Proj-0} pairwise, we obtain $P^{\mu\rho}_{\mu\rho}=\frac{1}{3}\theta^\mu_\mu \theta^\rho_\rho=3$, where each $\theta^\mu_\mu=3$. Recall that these projector are already derived for a perturbation around flat space, so the Newtonian limit does not impact their value. Substituting for $\Delta_2$ and 
$\Delta_w$ with $d=4$ and
substituting everything together, one obtains
\begin{equation}
\frac{1}{(2\pi)^3}\frac{P^{\mu \nu (0)}_{\mu \nu}}{\frac{1}{2}\Delta_w^{-1}+3\Delta_2^{-1}} = \frac{32 G^2_N }{\pi D_0(1-2\beta)}
\end{equation}
which will then multiply the result of the Fourier transform of $\frac{1}{p^4}$. 

In the case of the weak field limit, and scalar part of the metric, we write $h_{\mu\nu}=-2\Phi\delta_{\mu\nu}$ with $\Phi$ the Newtonian potential. If we now also contract the indices on the left hand side of Eq. \eqref{eq:hmunuTwoPoint}, and writing the Fourier transform of $\Phi(x)$ as $\tilde{\Phi}(p)$, we find

\begin{align}
    \langle \tilde{\Phi}(p) \tilde{\Phi}(p')\rangle-\langle \tilde{\Phi}{(p)}\rangle\langle \tilde{\Phi}(p')\rangle=\frac{2G_N^2}{\pi D_0(1-2\beta)}\frac{1}{p^4}\delta(p-p')
\end{align}
We now set the dimensionless coupling constant
\begin{align}
    D_G=\frac{2G_N^2}{\pi D_0(1-2\beta)}
    \label{eq:Dg}
\end{align}
Note that in \cite{layton2023weak,UCLDMDNE}, the quantity $\frac{G_N^2}{8\pi(1-\beta) D_0}$ sits in front of the non-relativistic action $I=- \int d^3x (\nabla^2\Phi)^2$, which can arise from different metric conventions, and the order of taking the weak field limit vs computing the two-point function. Since $\beta$ is an unknown constant which we expect to be of order unity, the difference in conventions is not relevant.

\subsection{Advanced, retarded and Feynman prescription}
\label{ss:gfuncs}

The Advanced and Retarded Green's functions for the $\Box^2$ operator are given by 
\begin{align}
    G_R(x-x')=-\frac{1}{8\pi}\theta(t-t'-|\vec{x}-\vec{x}'|)
    \label{eq:retarded}
\end{align}and
\begin{align}
    G_A(x-x')=\frac{1}{8\pi}\theta(t-t'+|\vec{x}-\vec{x}'|)
    \label{eq:advanced}
\end{align}
where $\theta$ is the Heaviside step function.
To see this, we write out $\Box^2$ acting on a function $\psi$, with the boundary conditions $\dot{\psi}=0,\ddot{\psi}=0,\dddot{\psi}=0$ and $\psi\rightarrow 0$ as $|x| \rightarrow \infty$
\begin{align}
    (\nabla^2-\partial_t^2)(\nabla^2-\partial_t^2)\psi(\vec{x},t)=f(\vec{x},t)
\end{align}
we obtain by a Laplace transform,
\begin{align}
    (\nabla^2-\omega^2)(\nabla^2-\omega^2)\psi(\vec{x},\omega)=f(\vec{x},\omega)
\end{align}
We then follow this with a Fourier transform, obtaining
\begin{align}
    (-\vec{k}^2-\omega^2)^2\psi(\vec{k},\omega)=f(\vec{k},\omega)
\end{align}
leading to
\begin{align}
   G(\vec{k},p)= \frac{-1}{(\vec{k}^2+\omega^2)^2}
\end{align}
We now perform an inverse Fourier transform
\begin{align}
 (2\pi)^3 G(\vec{x},p)=\int_{0}^\infty \int_{0}^{2\pi} \int_{0}^{\pi}\frac{-e^{ik|x|\cos\theta}}{(k^2+\omega^2)^2}k^2dk d\theta d\phi
\end{align}
Thus,
\begin{align}
    G(\vec{x},\omega)=\frac{-1}{(2\pi)^2}\int_{-\infty}^{\infty}\frac{e^{ik|x|}}{(k^2+\omega^2)^2}\frac{k}{i|x|}dk
\end{align}
We have poles of order 2 at $k=\pm i\omega$. Closing the contour in the upper plane, we obtain, 
\begin{align}
    G(\vec{x},\omega)=\frac{-e^{-\omega|x|}}{8\pi}
\end{align}
Performing the inverse Laplace transform gives us the retarded Green's function
\begin{equation}
    \label{eq:Inverse of box2}
    G(\vec{x},t)=\frac{-1}{8\pi}\theta(t-|x|)
\end{equation}
It is then straightforward to modify this to get 
Eq. \eqref{eq:retarded}, and Eq. \eqref{eq:advanced} for the advanced Green's function. Note that one can also construct Green's function, by convoluting a Green's function $G_1(x-x')$ for the d'Alembertian with itself. 
\begin{align}
    G_{2}(x-x')=\int d^4y\, G_1(x-y)G_1(x'-y).
    \label{eq:BoxsquaredGF} 
\end{align}

In order to obtain the advanced and retarded spectral densities we 
return to Eq. \eqref{eq:spectdensity}. Using the fact that Eq. \eqref{eq:spectralAtx}  is an even function of $k$, and using $x$ and $|x-x'|$ interchangeably in this section, we write
\begin{equation}
  \int^{\infty}_{-\infty} dk k^3 \frac{\sin{kx}}{(k^2-\omega^2)^2} = \frac{1}{2\pi i}\int \frac{(e^{ikx}-e^{-ikx}) k^3 dk}{(k^2-\omega^2)^2}
\end{equation}
For the retarded propagator, we
must have both poles below the real axis in the $\omega$-plane. Thus, we compute
\begin{equation}
   \frac{1}{2\pi i} \int^{\infty}_{-\infty} dk k^3 \frac{(e^{ikx}-e^{-ikx})}{(k-\omega+i\epsilon)^2(k+\omega-i\epsilon)^2}
\end{equation}

We have poles of order 2 at $k=\mp \omega \pm i\epsilon$. 
We integrate the $e^{ikx}$ part over a semicircle such that $Im(k)>0$. The
integral over the semi-circle vanishes and hence the integral over real axis is equal to $2\pi i$ times the residue.   
The residue is given by (suppressing the $i\epsilon$):
\begin{equation}
    Res_{k=-\omega}f(z)=\frac{d}{dk}((k+\omega)^2f(z))
\end{equation}
where 
\begin{equation}
    f(z) = \frac{(e^{ikx}) k^3 dk}{(k+\omega)^2(k-\omega)^2}
\end{equation}
We repeat this with $e^{-ikx}$ this time such that $Im(k)<0$, using a clockwise contour. We then add $2\pi Res_{k=-\omega}-2\pi Res_{k=\omega}$  to get
\begin{equation}
   \int^{\infty}_{-\infty} dk k^3 \frac{\sin{kx}}{(k^2-\omega^2)^2}=\pi\frac{2-i\omega x}{2}e^{-ix\omega}
   \label{eq:advancedSD}
\end{equation}
where $i\epsilon$ have been suppressed.
Using a similar procedure for the retarded Green's function, we obtain  \begin{equation}
    \int^{\infty}_{-\infty} dk k^3 \frac{\sin{kx}}{(k-\omega-i\epsilon)^2(k+\omega+i\epsilon)^2}= \pi \frac{2+i \omega x}{2} e^{i \omega x}.
    \label{eq:retardedSD}
\end{equation}


The Feynmann prescription meanwhile, gives us
\begin{equation}
    \int^{\infty}_{-\infty} dk k^3 \frac{\sin{kx}}{(k-\omega-i\epsilon)^2(k+\omega-i\epsilon)^2}=\pi \left(\cos{\omega x} -\frac{1}{2}\omega x \sin{\omega x}\right)
    \label{eq:someieps}
\end{equation}
Note that if we change both signs of $i \epsilon$ terms in the above integral we obtain the same result.

Putting these three integrals into Eq. \eqref{eq:spectralAtx}, we present the spectral densities. For the retarded function, we obtain 
\begin{equation}
   S_{aa,R}(\vec{x},\vec{x}',\omega)= \frac{D_G}{ 4\pi } \left(\frac{1}{|x-x'|}+\frac{i \omega}{2}\right) e^{i \omega |x-x'|}.
  \label{eq:spectralAtx2}
\end{equation}
For the advanced propagator we obtain
\begin{equation}
 S_{aa,A}(\vec{x},\vec{x}',\omega)= \frac{D_G}{ 4\pi }\left( \frac{1}{|x-x'|}-\frac{i \omega}{2} \right)e^{-i \omega |x-x'|}
\label{eq:spectralRtx2}
\end{equation}.
Note that for stationary test masses, only the $1/|x-x'|$ contribution is relevant, since the physically relevant quantity is Eq. \eqref{eq:spectralAtx2}  and \eqref{eq:spectralRtx2} integrated over a test mass, and the $i\omega$ term has a Fourier transform which goes as $\delta'(t-t'\pm |x-x'|$ and vanishes after integration by parts.

%

\subsection{Schwinger-Keldysh prescription}

\label{ss:SK}



The cut propagator, which propagates between bra and ket fields is given by the imaginary part of the Feynmann propagator: 
$\text{Im}(\Delta(p))=\frac{1}{2i}\left(\frac{1}{p^2+m^2-i\epsilon}-\frac{1}{p^2+m^2+i\epsilon}\right)$, which amounts to putting the intermediate particles onshell: $\text{Im}(\Delta(p))\rightarrow \delta(p^2+m^2)$. The mod-squared prescription for $\frac{1}{p^4}$, on the other hand, can also be written as $\frac{1}{\epsilon}\text{Im}\Delta$ for a massless particle.

\subsection{Massive case}

For completeness, we also consider the massive case, as one could be induced when the matter contribution is added to the action, given that it corresponds to an $R$ term in the action. It also cures the IR divergence. In the case where $\alpha\neq 0$, and the propagator is of the form of Equation \eqref{eq:twopoint}, this can be achieved by writing the propagator in terms of partial fractions as in Eq. \eqref{eq:partialfracs}. We will instead consider  the spectral density corresponding to the propagator
\begin{align}
    D(p)= \frac{1}{(p^2+m^2)^2}
\end{align}
and assuming $\omega^2 \geq m^2$ we obtain (where we have suppressed the respective Heaviside functions)
\begin{equation}
    \int^{\infty}_{-\infty} dk k^3 \frac{\sin{kx}}{(\sqrt{k^2+m^2}-\omega-i\epsilon)^2(\sqrt{k^2+m^2}+\omega+i\epsilon)^2}=2 \pi \frac{2+i \sqrt{\omega^2-m^2}x}{4} e^{i \sqrt{\omega^2-m^2}x}.
    \label{eq:retardedSD}
\end{equation}
for the retarded Green function and
\begin{equation}
    \int^{\infty}_{-\infty} dk k^3 \frac{\sin{kx}}{(\sqrt{k^2+m^2}-\omega+i\epsilon)^2(\sqrt{k^2+m^2}+\omega-i\epsilon)^2}=2 \pi \frac{2-i \sqrt{\omega^2-m^2}x}{4} e^{-i \sqrt{\omega^2-m^2}x}
    \label{eq:retardedSD}
\end{equation}
for the advanced one.
For $\omega^2 < m^2$ we obtain for retarded and advanced Green functions the same expression
\begin{eqnarray}
    \int^{\infty}_{-\infty} dk k^3 \frac{\sin{kx}}{(\sqrt{k^2+m^2}-\omega-i\epsilon)^2(\sqrt{k^2+m^2}+\omega+i\epsilon)^2}=\nonumber\\ \int^{\infty}_{-\infty} dk k^3 \frac{\sin{kx}}{(\sqrt{k^2+m^2}-\omega+i\epsilon)^2(\sqrt{k^2+m^2}+\omega-i\epsilon)^2}=\nonumber\\
    2 \pi \frac{2-\sqrt{m^2-\omega^2}x}{4} e^{-\sqrt{m^2-\omega^2}x}.
    \label{eq:retardedSD}
\end{eqnarray}


The Fourier transform for the time-like case is given by

\begin{align}
    G(x-x',t-t') =[\frac{1}{2i\epsilon}]\large[ \frac{\sqrt{m^2-i\epsilon}}{4\pi^2s}K_1\left(is\sqrt{m^2-i\epsilon}\right)-\frac{\sqrt{m^2+i\epsilon}}{4\pi^2s}K_1\left(-is\sqrt{m^2+i\epsilon}\right)]
\end{align}
 For the case of
 \begin{align}
     G(k)=\frac{1}{p^2(p^2+\alpha^2)}=\frac{1}{(p^2+\frac{\alpha^2}{2})^2-\frac{\alpha^4}{4}}=\frac{1}{\alpha^2}\left(\frac{1}{p^2}-\frac{1}{p^2+\alpha^2}\right)
 \end{align}

this is positive semi-definite for all $(\vec{p})^2>\omega^2$.

In general, its spectral density is given by

\begin{align}
    S_{aa,S}=\frac{1}{\alpha^2|x-x'|8\pi i}(\omega_1^{2}e^{i\omega_1|x-x'|}-\omega_2^{2}e^{-i\omega_2|x-x'|})
\end{align}

where $\omega_1=\sqrt{\omega+i\epsilon}$ and $\omega_2=\sqrt{\omega^2-\alpha^2-i\epsilon}$

whereas, its Fourier transform is given by 

\begin{align}
    G(x-x',t-t') =[\frac{1}{\alpha^2}]\large[ \frac{\sqrt{-i\epsilon}}{4\pi^2s}K_1\left(is\sqrt{-i\epsilon}\right)-\frac{\sqrt{\alpha^2+i\epsilon}}{4\pi^2s}K_1\left(-is\sqrt{\alpha^2+i\epsilon}\right)]
\end{align}

Note that unlike both the massless and massive cases, this two-point function is not IR-divergent.

\section{Dimensional regularisation of the IR divergence}
\label{sec:dimreg}

Here, we show that the two-point function Eq.~\eqref{eq:nofreedom} 
\begin{align}
\label{eq:EGtwo}
    G_2(\vec{x}-\vec{x}')=\frac{1}{(4\pi)^2}\int\! d^3y \,\frac{1}{|\vec{y}-\vec{x}||\vec{y}-\vec{x}'|},
\end{align}
discussed in \cite{oppenheim2022gravitationally}, after dimensional regularisation, \textit{i.e.} on employing analytic continuation in the dimension $d=3-2\epsilon$, gives the two-point function of Eq. \eqref{eq:freedom},
\begin{align}
\label{eq:EG}
    G(\vec{x}-\vec{x}')=-\frac{1}{8 \pi}|\vec{x}-\vec{x}'|.
\end{align}
Thus we start from 
\begin{equation}
    \int\! d^{3-2\epsilon}y'\, \frac{1}{|\vec{y}\,'-\vec{x}||\vec{y}\,'-\vec{x}'|} = \int\! d^{3-2\epsilon}y\, \frac{1}{y|\vec{y}+\vec{z}|}
\end{equation}
where we shifted $\vec{y}{\,'}=\vec{y}+\vec{x}$ and set $\vec{z}=\vec{x}-\vec{x}'$. Combining denominators using Feynman parameters and using standard loop-integral formulae for the resulting integral:
\begin{equation}
\begin{split}
\int\! d^{3-2\epsilon}y\, \frac{1}{y|\vec{y}+\vec{z}|} &= \frac{\Gamma(1)}{\Gamma^2(\tfrac12)} \int d^{3-2\epsilon}y\int_0^1\!\! d\alpha_1 d\alpha_2\, \frac{\delta(1-\alpha_1-\alpha_2)\alpha_1^{-\frac{1}{2}}\alpha_2^{-\frac{1}{2}} }{\alpha_1y^2+\alpha_2(\vec{y}+\vec{z})^2} \\
&=\frac{1}{\pi}\int\! d^{3-2\epsilon}y\, \int_0^1\!\! d\alpha \,\frac{(1-\alpha)^{-\frac{1}{2}}\alpha^{-\frac{1}{2}}}{(\vec{y}+\alpha \vec{z})^2 + \alpha(1-\alpha)z^2} \\
&=\frac{1}{\pi} \frac{(2\pi)^{3-2\epsilon}}{(4\pi)^{\frac{3}{2}-\epsilon}}\Gamma\left(-\tfrac12+\epsilon\right)\int_0^1\!\! d\alpha \,\big[\alpha(1-\alpha)z^2\big]^{\frac{1}{2}-\epsilon}\alpha^{-\frac{1}{2}}(1-\alpha)^{-\frac{1}{2}}\\
&=(\pi z^2)^{1/2-\epsilon}\,\Gamma(-\tfrac12+\epsilon)  \int_0^1\!\! d\alpha\, \alpha^{-\epsilon}(1-\alpha)^{-\epsilon}\quad \to\quad -2\pi z\,,
\end{split}
\end{equation}
where in the last step we recognize that the expression has a finite limit as $\epsilon\to0$. This finite answer is the result of analytically continuing around the pole at $d=2$ ($\epsilon=\tfrac12$). Substituting the result into \eqref{eq:EGtwo} we get \eqref{eq:EG} as claimed.

Had we computed the variance of the acceleration directly we would avoid the infrared divergence:
\begin{equation}
\label{eq:Evaracc}
    \frac{\partial}{\partial x_i} \frac{\partial}{\partial x'_j} G_2(\vec{x}-\vec{x}') = -\frac{\partial^2}{\partial z_i \partial z_j} G_2(z) = \frac{1}{(4\pi)^2}\int\!d^3y\left\{ \frac{\delta_{ij}}{y |\vec{y}+\vec{z}|^3}-3\frac{(y+z)_i (y+z)_j}{y|\vec{y}+\vec{z}|^5}\right\}
\end{equation}
but replace it with an ultraviolet divergence (seen here in the limit $\vec{y}\to-\vec{z}$). Despite this, from above we expect the result to be finite in dimensional regularisation and be what we would obtain by using \eqref{eq:EG}:
\begin{equation}
\label{eq:Evaraccanswer}
  \frac{\partial}{\partial x_i} \frac{\partial}{\partial x'_j} G_2(\vec{x}-\vec{x}') =  -\frac{\partial^2}{\partial z_i \partial z_j} G(z) = \frac{1}{8\pi z}\left(\delta_{ij}-\frac{z_iz_j}{z^2}\right)\,.
\end{equation}
Using Feynman parametrisation, \eqref{eq:Evaracc} can be written as
\begin{equation}
   \frac{\delta_{ij}}{8\pi^3} \int\! d^{3-2\epsilon}y\, \int_0^1\!\! d\alpha \,
   \frac{(1-\alpha)^{-\frac{1}{2}}\alpha^{\frac{1}{2}}}{\left[(\vec{y}+\alpha \vec{z})^2 + \alpha(1-\alpha)z^2\right]^2}
   -\frac{1}{2\pi^3}\int\! d^{3-2\epsilon}y\, \int_0^1\!\! d\alpha \,
   \frac{(1-\alpha)^{-\frac{1}{2}}\alpha^{\frac{3}{2}}(y+z)_i(y+z)_j}{\left[(\vec{y}+\alpha \vec{z})^2 + \alpha(1-\alpha)z^2\right]^3}\,,
\end{equation}
This can again be tackled by standard loop-integral formulae, and indeed the result verifies \eqref{eq:Evaraccanswer}.

\section{Most probable path}
\label{app: mpp}

In this Appendix, we calculate the contribution from the matter action to the equations for the most probable path. We keep the original coupling constants of \cite{oppenheim2023covariant}, but with hindsight, one should allow each term be be free. To insure that pure quantum states evolve to pure quantum states conditioned on the spacetime, one should set all terms with interactions between $\phi^+$ and $\phi^-$ to zero. This is the equivalent of saturating the decoherence-vs-diffusion trade-off. We expect this condition is likely preserved under renormalisation, because the action then breaks up into one action for the $\phi^+$ field and one for the $\phi^-$ field.
We here neglect the normalisation term. We have the following contribution from the gravitational action (without the Gauss-Bonnet term)

\begin{equation}
\begin{split}
& -\frac{D_0}{128 \pi^2 G_{N}^2}\bigg[\Lambda(2-d)(1-\beta d)\left(R^{\kappa \lambda}-\frac{1}{2}g^{\kappa \lambda}R\right)+2\left(\nabla^{\alpha}\nabla^{\beta}+\frac{1}{2}R^{\alpha \beta}\right)C^{\kappa}\text{}_{\alpha}\text{}^{\lambda}\text{}_{\beta}+\nonumber\\ 
&\left(\frac{d}{4}-\frac{2}{3}-\beta \left(1-\frac{d}{2}\right)^2\right)\left(R^{\kappa \lambda}-\frac{1}{4}R g^{\kappa \lambda}-\nabla^{\kappa}\nabla^{\lambda}+g^{\kappa \lambda}\nabla^2\right)R-\frac{d\Lambda^2 (1 -\beta d)}{2}g^{\kappa \lambda}\bigg].
\end{split}
\end{equation}

Let us calculate the variation of the first term of the matter action in Equation~\eqref{eq:matteraction}

\begin{equation}
\label{eq: eqvar1}
-\frac{D_0}{16 \pi G_{N}}\int dx \, \left[\delta \sqrt{-g}\,(-2T^{\alpha \beta})R_{\alpha \beta}+\sqrt{-g}\left(\frac{\partial (-2T^{\alpha \beta})}{\partial g_{\kappa \lambda}}R_{\alpha \beta}\,\delta g_{\kappa \lambda}+(-2T^{\alpha \beta})\,\delta R_{\alpha \beta}\right)\right].  
\end{equation}

The variation of the square root of the metric determinant is given by

\begin{eqnarray}
\delta \sqrt{-g}=\frac{1}{2}\sqrt{-g}g^{\kappa \lambda}\delta g_{\kappa \lambda}.
\end{eqnarray}

Therefore, the variation of the first term results in

\begin{eqnarray}
-\frac{D_0}{16 \pi G_{N}}\int dx\, \left(\frac{1}{2}\sqrt{-g}g^{\kappa \lambda}\delta g_{\kappa \lambda}(-2T^{\alpha \beta})R_{\alpha \beta}\right).  
\end{eqnarray}

The last term is of the form $\int d x F^{\alpha \beta} \delta R_{\alpha \beta}$, containing the variation of the Ricci tensor. The variation of terms of such a form is

\begin{eqnarray}
\int dx\, F^{\alpha \beta} \delta R_{\alpha \beta}=\int dx\, F^{\alpha \beta}\big( (\delta \Gamma^{\mu}_{\alpha \beta})_{;\mu}-(\delta \Gamma^{\mu}_{\alpha \mu})_{;\beta}\big).  
\end{eqnarray}

We simplify this further integrating by parts and renaming some indices to get to

\begin{eqnarray}
\int dx\, (\delta \Gamma^{\mu}_{\alpha \beta})_{;\mu}F^{\alpha \beta}=\int dx\, \big(\delta \Gamma^{\mu}_{\alpha \beta}(-F^{\alpha \beta}_{,\mu}+\Gamma^{\gamma}_{\gamma \mu} F^{\alpha \beta}- \Gamma^{\alpha}_{\mu \gamma}F^{\gamma \beta}- \Gamma^{\beta}_{\mu \gamma}F^{\alpha \gamma})\big)=\int dx\,\delta \Gamma^{\mu}_{\alpha \beta}G_{\mu}^{\alpha \beta}(F^{\gamma \delta}),    
\end{eqnarray}
for the first term and to

\begin{eqnarray}
\int dx\, (\delta \Gamma^{\mu}_{\alpha \mu})_{;\beta}F^{\alpha \beta}=\int dx\, \big(\delta \Gamma^{\mu}_{\alpha \mu}(-F^{\alpha \beta}_{,\beta}- \Gamma^{\alpha}_{\beta \gamma}F^{\gamma \beta})\big)=\int dx\, \delta \Gamma^{\mu}_{\alpha \mu}H^{\alpha}(F^{\gamma \delta}),     
\end{eqnarray}
for the second.
Expanding the variation of the Christoffel symbol, we can write

\begin{eqnarray}
\label{eq: eqvar8}
\int dx\, \delta \Gamma^{\mu}_{\alpha \beta}G^{\alpha \beta}_{\mu}=\int dx\, \left(\frac{1}{2}\delta g^{\mu \delta}(\partial_{\alpha}g_{\beta \delta}+\partial_{\beta}g_{\alpha \delta}-\partial_{\delta}g_{\alpha \beta})+\frac{1}{2} g^{\mu \delta}(\partial_{\alpha}\delta g_{\beta \delta}+\partial_{\beta}\delta g_{\alpha \delta}-\partial_{\delta}\delta g_{\alpha \beta})\right)G^{\alpha \beta}_{\mu},
\end{eqnarray}

\begin{equation}
\label{eq: eqvar9}
\begin{split}
\int dx\, \delta \Gamma^{\mu}_{\alpha \mu}H^{\alpha}& =\int dx\, \left(\frac{1}{2}\delta g^{\mu \delta}(\partial_{\alpha}g_{\mu \delta}+\partial_{\mu}g_{\alpha \delta}-\partial_{\delta}g_{\alpha \mu})+\frac{1}{2} g^{\mu \delta}(\partial_{\alpha}\delta g_{\mu \delta}+\partial_{\mu}\delta g_{\alpha \delta}-\partial_{\delta}\delta g_{\alpha \mu})\right)H^{\alpha}\\
&= \int dx\, \left(\frac{1}{2}\delta g^{\mu \delta}(\partial_{\alpha}g_{\mu \delta})+\frac{1}{2} g^{\mu \delta}(\partial_{\alpha}\delta g_{\mu \delta})\right)H^{\alpha},
\end{split}
\end{equation}
where in the last equality we assumed that variations are symmetric.

The variation of $g^{\alpha \beta}$ is

\begin{eqnarray}
\delta g^{\alpha \beta}=-g^{\alpha \kappa}g^{\beta \lambda}\delta g_{\kappa \lambda}.
\end{eqnarray}

Substituting back in Eqs.~\eqref{eq: eqvar8} and~\eqref{eq: eqvar9}  we obtain

\begin{align}
\int dx\, \delta \Gamma^{\mu}_{\alpha \beta}G^{\alpha \beta}_{\mu} &=\int dx\, \bigg[\frac{1}{2}g^{\mu \kappa} g^{\delta \lambda}(\partial_{\alpha}g_{\beta \delta}+\partial_{\beta}g_{\alpha \delta}-\partial_{\delta}g_{\alpha \beta})G^{\alpha \beta}_{\mu}-\partial_{\alpha}\left(\frac{1}{2} g^{\mu \lambda}G^{\alpha \kappa}_{\mu}\right) \nonumber\\
&\quad\quad\quad\quad\quad -\partial_{\beta}\left(\frac{1}{2} g^{\mu \lambda}G^{\kappa \beta}_{\mu}\right)  +\partial_{\delta}\left(\frac{1}{2} g^{\mu \delta}G^{\kappa \lambda}_{\mu}\right)\bigg] \delta g_{\kappa \lambda},\\
\int  dx\, \delta \Gamma^{\mu}_{\alpha \mu}H^{\alpha}&=\int dx\, \left[\frac{1}{2}g^{\mu \kappa} g^{\delta \lambda}(\partial_{\alpha}g_{\mu \delta})H^{\alpha}-\partial_{\alpha}\left(\frac{1}{2} g^{\kappa \lambda}H^{\alpha}\right)\right] \delta g_{\kappa \lambda}.
\end{align}

At this point, we are quite close to our goal. We can now write

\begin{equation}
\int  dx\, F^{\alpha \beta} \delta R_{\alpha \beta}= \int  dx\, I^{\kappa \lambda}(F^{\gamma \delta})\delta g_{\kappa \lambda},
\end{equation}
where

\begin{equation}
\begin{split}
I^{\kappa \lambda}(F^{\gamma \delta}) &= \bigg[\frac{1}{2}g^{\mu \kappa} g^{\delta \lambda}(\partial_{\alpha}g_{\beta \delta}+\partial_{\beta}g_{\alpha \delta}-\partial_{\delta}g_{\alpha \beta})G^{\alpha \beta}_{\mu}-\partial_{\alpha}\left(\frac{1}{2} g^{\mu \lambda}G^{\alpha \kappa}_{\mu}\right) -\partial_{\beta}\left(\frac{1}{2} g^{\mu \lambda}G^{\kappa \beta}_{\mu}\right)  +\partial_{\delta}\left(\frac{1}{2} g^{\mu \delta}G^{\kappa \lambda}_{\mu}\right)\\
& -\frac{1}{2}g^{\mu \kappa} g^{\delta \lambda}(\partial_{\alpha}g_{\mu \delta})H^{\alpha}+\partial_{\alpha}\left(\frac{1}{2} g^{\kappa \lambda}H^{\alpha}\right)\bigg].
\end{split}
\end{equation}

Taking these expressions into account, we have

\begin{eqnarray}
\int  dx\, \sqrt{-g}(-2T^{\alpha \beta})\delta R_{\alpha \beta} =\int  dx\, I^{\kappa \lambda}(-2T^{\gamma \delta}))\delta g_{\kappa \lambda}
\end{eqnarray}

We can now substitute back in the variation given by Equation~\eqref{eq: eqvar1} to see that the three contributions to the equation for most probable path are given by

\begin{equation}
-\frac{D_0}{16 \pi G_{N}}\left[\frac{1}{2}\sqrt{-g}g^{\kappa \lambda}(-2T^{\alpha \beta})R_{\alpha \beta}+\sqrt{-g}\frac{\partial (-2T^{\alpha \beta})}{\partial g_{\kappa \lambda}}R_{\alpha \beta}+I^{\kappa \lambda}\big(\sqrt{-g}(-2T^{\gamma \delta})\big)\right].
\end{equation}

Let us now calculate the variation of the second term of the matter action

\begin{equation}
\label{eq: eqvar2}
-\frac{D_0}{16 \pi G_{N}}\big(1+\beta(2-d)\big)\int  dx\, \left[\delta \sqrt{-g}\,\bar{T}R+\sqrt{-g}\left(\frac{\partial \bar{T}}{\partial g_{\kappa \lambda}}R\,\delta g_{\kappa \lambda}+\bar{T}\delta g^{\alpha \beta} R_{\alpha \beta}+\bar{T} g^{\alpha\beta}\,\delta R_{\alpha \beta}\right)\right].
\end{equation}

The variation of the third term in the above equation is

\begin{eqnarray}
\label{eqvar2}
-\frac{D_0}{16 \pi G_{N}}\big(1+\beta(2-d)\big)\int dx\, \sqrt{-g}\,\bar{T}(-g^{\alpha \kappa}g^{\beta \lambda}\delta g_{\kappa \lambda}) R_{\alpha \beta}.
\end{eqnarray}

Taking into account the variations of the first term and the fourth term (which are the same as calculated before) we see that Equation~\eqref{eq: eqvar2} contributes the following terms to the equation for the most probable path

\begin{equation}
-\frac{D_0}{16 \pi G_{N}}\big(1+\beta(2-d)\big) \left[\frac{1}{2}\sqrt{-g}\bar{T}R\, g^{\kappa \lambda}+\sqrt{-g}\frac{\partial \bar{T}}{\partial g_{\kappa \lambda}}R+\sqrt{-g}\bar{T}(-g^{\alpha \kappa}g^{\beta \lambda}) R_{\alpha \beta}+I^{\kappa \lambda}(\sqrt{-g}\;\bar{T} g^{\gamma \delta})\right].
\end{equation}

Let us calculate the variation of the last three terms

\begin{equation}
\begin{split}
-\frac{D_0}{16 \pi G_{N}}\int dx\, \bigg[&\delta \sqrt{-g}(-8 \pi G_{N}\beta\bar{T}^2+8 \pi G_{N}\bar{T}^{\alpha \beta}\bar{T}_{\alpha \beta}+2(d\beta-1)\Lambda\bar{T}\\
& +\sqrt{-g}\frac{\partial (-8 \pi G_{N}\beta\bar{T}^2+8 \pi G_{N}\bar{T}^{\alpha \beta}\bar{T}_{\alpha \beta}+2(d\beta-1)\Lambda\bar{T}}{\partial g_{\kappa \lambda}}\delta g_{\kappa \lambda}\bigg].
\end{split}
\end{equation}

Hence, the contribution from this term to the most probable path equation is:

\begin{equation}
\begin{split}
-\frac{D_0}{16 \pi G_{N}} \bigg[& \sqrt{-g}g^{\kappa \lambda}(-8 \pi G_{N}\beta\bar{T}^2+8 \pi G_{N}\bar{T}^{\alpha \beta}\bar{T}_{\alpha \beta}+2(d\beta-1)\Lambda\bar{T}\\
& +\sqrt{-g}\frac{\partial (-8 \pi G_{N}\beta\bar{T}^2+8 \pi G_{N}\bar{T}^{\alpha \beta}\bar{T}_{\alpha \beta}+2(d\beta-1)\Lambda\bar{T}}{\partial g_{\kappa \lambda}}\bigg].
\end{split}
\end{equation}

Taking all expressions into account, we lastly obtain the following equation that determines the most probable path 

\begin{equation}
\label{eq: gravity_mpp}
\begin{split}
&\frac{1}{128 \pi^2 G_{N}^2}\left(\Lambda(2-d)(1-\beta d)\left(R^{\kappa \lambda}-\frac{1}{2}g^{\kappa \lambda}R\right)+2\left(\nabla^{\alpha}\nabla^{\beta}+\frac{1}{2}R^{\alpha \beta}\right)C^{\kappa}\text{}_{\alpha}\text{}^{\lambda}\text{}_{\beta} \right.\\
&\left.+\left(\frac{d}{4}-\frac{2}{3}-\beta \left(1-\frac{d}{2}\right)^2\right)\left(R^{\kappa \lambda}-\frac{1}{4}R g^{\kappa \lambda}-\nabla^{\kappa}\nabla^{\lambda}+g^{\kappa \lambda}\nabla^2\right)R-\frac{d\Lambda^2 (1 -\beta d)}{2}g^{\kappa \lambda}\right) \\
& +\frac{1}{16 \pi G_{N}}\left(\frac{1}{2}\sqrt{-g}g^{\kappa \lambda}(-2T^{\alpha \beta})R_{\alpha \beta}+\sqrt{-g}\frac{\partial (-2T^{\alpha \beta})}{\partial g_{\kappa \lambda}}R_{\alpha \beta}+I^{\kappa \lambda}(\sqrt{-g}(-2T^{\gamma \delta}))\right)\\
& +\frac{1}{16 \pi G_{N}}\big(1+\beta(2-d)\big)\left(\frac{1}{2}\sqrt{-g}g^{\kappa \lambda}\bar{T}R+\sqrt{-g}\frac{\partial \bar{T}}{\partial g_{\kappa \lambda}}R+\sqrt{-g}\bar{T}(-g^{\alpha \kappa}g^{\beta \lambda}) R_{\alpha \beta}+I^{\kappa \lambda}(\sqrt{-g}\bar{T} g^{\gamma \delta})\right)\\
& +\frac{1}{16 \pi G_{N}}\bigg(\frac{1}{2}\sqrt{-g}g^{\kappa \lambda}(-8 \pi G_{N}\beta\bar{T}^2+8 \pi G_{N}\bar{T}^{\alpha \beta}\bar{T}_{\alpha \beta}+2(d\beta-1)\Lambda\bar{T}\\
& -\sqrt{-g}\frac{\partial (-8 \pi G_{N}\beta\bar{T}^2+8 \pi G_{N}\bar{T}^{\alpha \beta}\bar{T}_{\alpha \beta}+2(d\beta-1)\Lambda\bar{T}}{\partial g_{\kappa \lambda}}\bigg)=0.
\end{split}
\end{equation}

\section{Power-counting renormalisability of the path integral}
\label{sec:renorm}

Here we recall the proof of power-counting renormalisability. For a full proof of renormalisability in the context of quadratic gravity, see ~\cite{stelle1977renormalization}. 

Consider a Feynman diagram with $N_V$ vertices, $N_I$ internal lines, and $N_L$ loops. Each propagator contributes a factor $q^{-4}$, each vertex contributes a factor $q$ to the power up to $4$, and each loop contributes a factor $q^4$. Hence for most divergent diagram we have
\begin{eqnarray}
q^{D_N}=q^{4N_V}q^{4N_L}q^{-4N_I},   
\end{eqnarray}
where $D_N$ is superficial degree of divergence. Using the topological identity
\begin{eqnarray}
N_I=N_L+N_V-1
\end{eqnarray}
we can write
\begin{eqnarray}
q^{D_N}=q^{4N_V}q^{4N_L}q^{4(1-N_L-N_V)}=q^4.   
\end{eqnarray}
Hence the superficial degree of divergence does not increase with the number of loops and the theory is power counting-renormalisable.

\section{Gauge fixing the path integral}
\label{sec:gauge}
We constrain path integral
\begin{eqnarray}
\int{Dg e^{S(g)}}.
\end{eqnarray}
to integration over metrics satisfying the harmonic gauge condition
\begin{eqnarray}
\partial_{\nu}h^{\mu \nu}=0
\end{eqnarray}
where the metric density splits in the following way
\begin{eqnarray}
\sqrt{-g}g^{\mu \nu}=\eta^{\mu \nu}+\kappa h^{\mu \nu},
\end{eqnarray}
where $\kappa^2=32 \pi G$.
We use the standard method to write path integral in the form (see ~\cite{stelle1977renormalization})
\begin{eqnarray}
\int Dh^{\mu \nu} D\bar{C}_{\tau} DC^{\alpha} e^{{S(h)}+S_{GF}(h)+S_{gh}(\bar{C},C,h)}.   
\end{eqnarray}
\begin{eqnarray}
S_{GF}=-\frac{1}{2}\kappa^2\delta^{-1}\int dx F_{\tau} \Box^2 F^{\tau},    
\end{eqnarray}
is the gauge fixing action, where
\begin{eqnarray}
F_{\tau}=F^{\tau}_{\mu \nu}h^{\mu \nu},
\end{eqnarray}
\begin{eqnarray}
F^{\tau}_{\mu \nu}=\delta^{\tau}_{\mu} \partial_{\nu},  
\end{eqnarray}
and $\delta$ is positive constant going to $0$.
\begin{eqnarray}
S_{ghost}=-\int dx \bar{C}_{\tau} F^{\tau}_{\mu \nu} D^{\mu \nu}_{\alpha} C^{\alpha}.
\end{eqnarray}
is ghost action. $C^{\alpha}$ and $\bar{C}_{\tau}$ are ghost and anti-ghost fields.
\begin{eqnarray}
D^{\mu \nu}_{\alpha}\xi^\alpha=\partial^{\mu}\xi^{\nu}+\partial^{\nu}\xi^{\mu}-\eta^{\mu \nu}\partial_{\alpha}\xi^\alpha+\kappa(\partial_{\alpha}\xi^{\mu}h^{\alpha \nu}+\partial_{\alpha}\xi^{\nu}h^{\alpha \mu}-\xi^{\alpha}\partial_{\alpha}h^{\mu \nu}-\partial_{\alpha}\xi^{\alpha}h^{\mu \nu})
\end{eqnarray}
generates gauge transformation of $h^{\mu \nu}$, i.e. $\delta h^{\mu \nu}= D^{\mu \nu}_{\alpha}\xi^\alpha$, where $\xi^\alpha(x)$ is a vector field corresponding to coordinate transformation $x^{\mu'}=x^\mu+\kappa\xi^\mu$

\end{document}